\documentclass{aa}
\usepackage{txfonts}
\usepackage{graphicx}
\usepackage{natbib}
\usepackage{rotating}
\bibpunct{(}{)}{;}{a}{}{,} 

\newcommand{\micron}{\mbox{$\mu$m}}

\begin{document}

	\title{Wide field CO \textit{J} = 3$\rightarrow$2 mapping of the Serpens Cloud Core}
	\author{O. Dionatos\inst{1,2}
          \and
	  B. Nisini
	   \inst{2}
	  \and
	  C. Codella
	   \inst{3}
	  \and
	  T. Giannini
	   \inst{2}
          }

\institute{
		Centre for Star and Planet Formation, Natural History Museum of Denmark, University of Copenhagen, {\O}ster Voldgade 5-7 DK-1350 Copenhagen, Denmark \\
              \email{odysseas@snm.ku.dk}\\
              \and
		INAF, Osservatorio Astronomico di Roma, Via di Frascati, 33  00040 Monte Porzio Catone (RM), Italy\\
         \and
		INAF, Osservatorio Astrofisico di Arcetri, Largo E. Fermi 5, 50125 Firenze, Italy\\
             }


\abstract
{Outflows provide indirect means to get an insight on diverse star formation associated phenomena. On scales of individual protostellar cores, outflows combined with intrinsic core properties can be used to study the mass accretion/ejection process of heavily embedded protostellar sources.}  
{The main objective of the paper is to study the overall outflow
distribution and association with the young population of the Serpens
Core cluster. In addition, the paper addresses the correlation of the
outflow momentum flux with the bolometric luminosity of their driving
sources using this homogeneous dataset for a single star forming site.}
{An area comprising 460$\arcsec\times$230$\arcsec$ of the Serpens cloud core has been mapped in $^{12}$CO \textit{J} = 3$\rightarrow$2 with the HARP-B heterodyne array at the James Clerk Maxwell Telescope;  \textit{J} = 3$\rightarrow$2 observations are more sensitive tracers of hot outflow gas than lower \textit{J} CO transitions; combined with the high sensitivity of the HARP-B receptors outflows are sharply outlined, enabling their association with individual protostellar cores. }
{Most of $\sim$20 observed outflows are found to be associated with known protostellar sources in bipolar or unipolar configurations. All but two outflow/core pairs in our sample tend to have a projected orientation spanning roughly NW-SE. The overall momentum driven by outflows in Serpens lies between 3.2 and 5.1 $\times$ 10$^{-1}$ M$_\odot$km s$^{-1}$, the kinetic energy from 4.3 to 6.7 $\times$ 10$^{43}$ erg  and momentum flux is between 2.8 and 4.4 $\times$ 10$^{-4}$ M$_\odot$ km s$^{-1}$ yr$^{-1}$.  Bolometric luminosities of protostellar cores based on Spitzer photometry are found up to an order of magnitude lower than previous estimations derived with IRAS/ISO data.   
}
{We confirm the validity of the existing correlations between the momentum flux and bolometric luminosity of Class I sources for the homogenous sample of Serpens, though we suggest that they should be revised by a shift to lower luminosities. All protostars classified as Class 0 sources stand well above the known Class I correlations, indicating a decline in momentum flux between the two classes.}

\keywords{stars: formation - sub-mm lines: ISM - ISM: jets and outflows - ISM: molecules}

\maketitle


\section {Introduction}

The Serpens cloud core is an active region of low mass star formation harboring a great diversity of star-birth associated phenomena. Since its first identification as a stellar nursery by \citet{Strom}, several studies have revealed several mm sources identified as protostellar condensations \citep{Testi1}, $\sim$10 Class 0 candidates  \citep{Casali, Hurt1, Harvey, Winston},  at least 20 Class I sources \citep{Kaas2, Winston, Harvey} along with a more evolved population of $\sim$ 150 Class II sources detectable in the near-IR \citep{Eiroa3, Giovannetti,  Kaas1}. The remarkably high stellar density and star formation efficiency  \citep[$\sim$5$\%$][]{Evans}, along with its proximity \citep[$\sim$ 310 pc][]{deLara} have rendered Serpens an observationally attractive site for star formation studies, as evidenced by the extensive literature.  

Star formation activity in Serpens is also manifested by dynamic mass-loss phenomena such as optical and near-IR detections of Herbig Haro objects \citep{Gomez, Reipurth, Davis, Ziener},  near-IR molecular jets \citep{Hodapp, Eiroa2, Herbst} and molecular outflows. Individual outflows in Serpens have been recorded in a wide variety of molecular tracers and different transitions (e,g,  CS \citet{Testi2, Mangum, Wolf-Chase}, H$_2$CO \citet{Mangum}, CH$_3$OH \citet{Garay, Testi2}, HCO$^+$ \citet{Gregersen, Hogerheijde}, HCN, SO and SiO \citet{Hogerheijde, Garay}, NH$_3$ \citep{Torrelles}). Wide field, homogenous outflow mapping of the Serpens Core has so far been performed only at the \textit{J} = 1$\rightarrow$0 \citep{Narayanan} and \textit{J} = 2$\rightarrow$1 \citep{White, Davis} transitions of carbon monoxide. This region has been mapped very recently also in the CO \textit{J} = 3$\rightarrow$2 as part of the JCMT Gould Belt Legacy Survey (GBS) to investigate the large scale mass and energetics of the cloud \citep{Graves}.

In this paper, we present a new CO \textit{J}=3$\rightarrow$2 map covering 460$\arcsec \times$ 230 $\arcsec$ 
of the Serpens core, with the aim of studying with an unprecedented detail the individual
outflows of the region; compared to lower \textit{J} maps, the higher energy transition (corresponding to a smaller beam for the same dish aperture) coupled with the higher sensitivity reached by the HARP-B observations delineate better than before individual outflows.  The advantage of higher energy transitions in mapping outflows is instructively illustrated in Figure 3 of \citet{Narayanan}, where higher-\textit{J} CO close-ups are presented towards the Class 0 source SMM4: higher \textit{J} maps suffer from much less confusion as the "warm" outflowing gas is better disentangled from the "cold" cloud material; this is particularly important in Serpens, where protostellar sources are closely clustered and their outflows tend to overlap and mix. 

Discriminating individual outflows enables the detailed investigation of a series of phenomena related to their generation. Outflow emission traces the swept-up ambient gas that has been entrained by a jet/wind that is directly linked to a protostar; in the opposite sense, i.e. backtracking the outflow lobes, the latter can be associated with individual protostellar sources.  In addition, outflow activity is thought to begin almost simultaneously with the formation of a hydrostatic core; therefore outflows paired to protostellar clumps can be employed as independent means for the identification of the existence of embedded protostars \citep[e.g.][]{Hatchell3}. 

Alternatively, outflows can be employed to study the accretion and mass ejection process in otherwise highly obscured and  unaccessible protostellar environments. There is increasing evidence that the momentum flux or "thrust" of molecular outflows is linked with the luminosity of the protostellar core \citep[e.g.][]{Cabrit2, Bontemps, Wu, Hatchell2}. Such correlation implies that independently of the details of the mass accretion and ejection mechanisms, both quantities are energetically governed in common by infall. Still, observational determination of both outflow thrust and bolometric luminosity in clustered protostellar environments remains challenging, and therefore their exact correlation remains not well defined.

In this context, in \S \ref{observations} we present the HARP-B observations and give an outline of the data reduction. Morphological characteristics of outflows and their association with individual protostellar sources are discussed in \S \ref{morphology}. In \S \ref{properties}, we derive outflow properties such as the column density, mass , momentum, kinetic energy and the momentum flux.. In \S \ref{luminosity} we estimate the bolometric luminosity employing recent Sptizer mid and far-IR photometry for the sources associated with the observed outflows. \S \ref{discussion} examines the correlation between the bolometric luminosity and momentrum flux, for the homogenous sample of sources in Serpens. In the same section, the relation between the bolometric luminosity with the outflow mass and mechanical luminosity are discussed. Finally, in \S \ref{conclusions} we summarize and draw the main conclusions of this work.  

\section {Observations and data reduction}
\label{observations}
$^{12}$CO \textit{J} = 3$\rightarrow$2 on-the-fly (OTF) maps were obtained between 09 April and 09 June 2007 with the HARP-B heterodyne array \citep{Smith} and ACSIS correlator \citep{Dent, Lightfoot} on the James Clerk Maxwell Telescope (JCMT). The mapped area was covered by consecutive scans in "basket-weave" mode at a position angle of 322$^o$. Each scan was offset from its neighbourings by 87.3$\arcsec$ in the orthogonal direction and signal was integrated every 7.3$\arcsec$ along the scan direction with an effective integration time of 3 seconds per point.  The resulting map dimensions are 7.7$\arcmin\times$4.3$\arcmin$, however the fully sampled area is somewhat smaller as the borders are not fully sampled due to a number of non-functional HARP-B receptors at the time the observations were performed. The spectral resolution is 488 KHz which corresponds to 0.41 km s$^{-1}$ at the frequency of 345.795 GHz. Single maps were coadded and initial data cubes were converted with makecube command in the STARLINK/SMURF package. The resulting map is centered at  \textit{$\alpha$} = 18$^h$29$^m$49$^s$.7, \textit{$\delta$} = 01$^o$15$\arcmin$22$\arcsec$.8 (epoch J2000) and has a pixel size of 7.3$\arcsec$, which is just below the Nyquist sampling limit for the $\sim$14$\arcsec$ beamwidth of each HARP-B receptor. A correction of n$_{MB}$=0.6 for the main beam efficiency was applied, and further reduction was performed with the CLASS/GreG packages. The mean \textit{rms} noise in brightness temperature T$_B$ per spectrum across the map is 0.08K, which is slightly deeper than the 0.1K reported in \citep{Graves}, while the 1-sigma \textit{rms} level for the map is at 0.27 K km s$^{-1}$.  Pixels exhibiting an \textit{rms} noise greater than 6-sigma of the mean were discarded in the final maps.

\section {Outflow Morphology}\label{morphology}
\subsection{Global}

\begin{figure*}
\centering
\includegraphics[width=17cm]{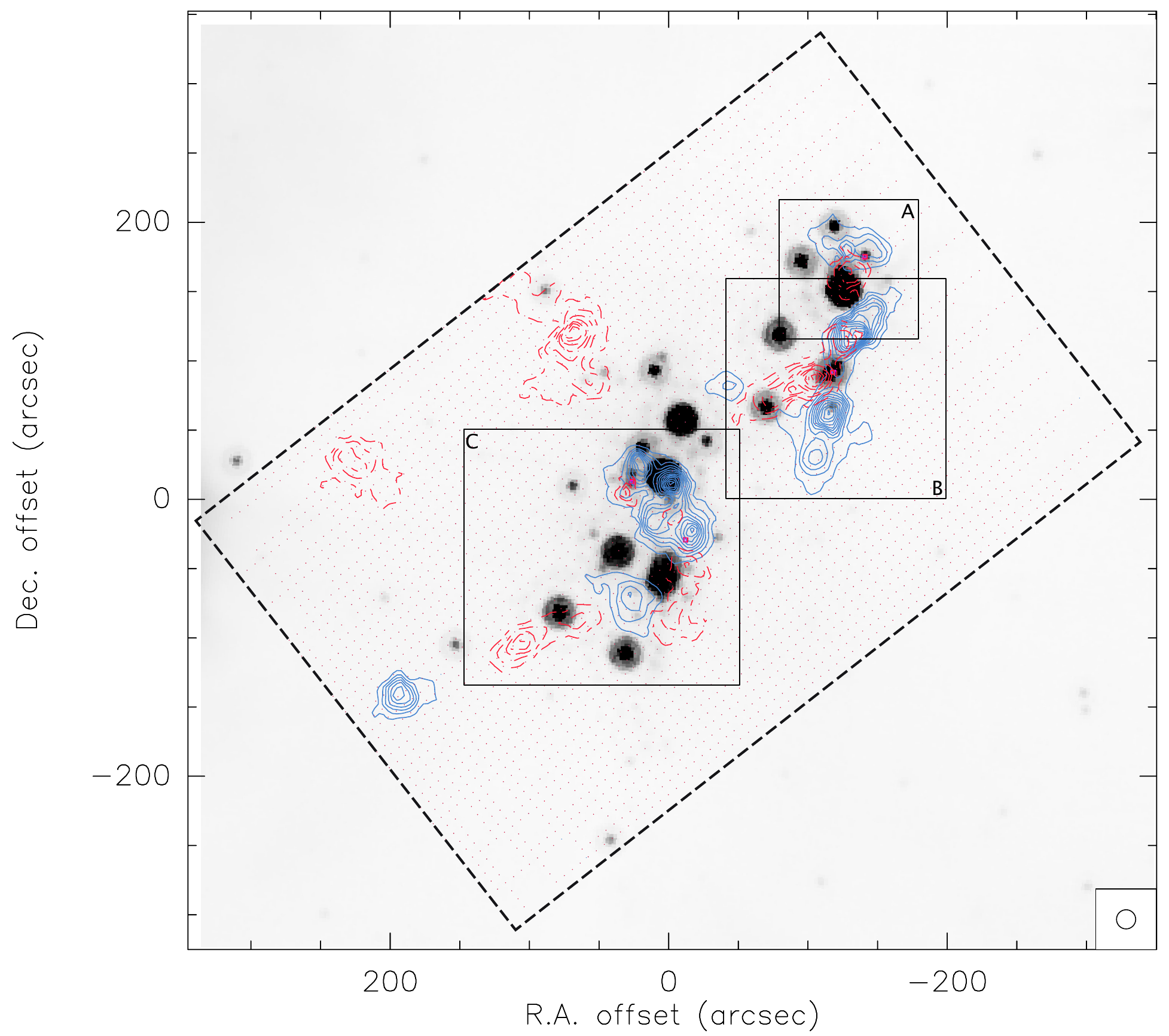}
\caption{High velocity CO \textit{J} = 3$\rightarrow$2 outflow map superimpsed on an MIPS band 1 (centered at 24 $\mu$m) image of the Serpens cloud core. Solid (blue) contours delineate blueshifted gas (integrated over -30 km s$^{-1}$ $\leqslant$ V$_{LSR}$ $\leqslant$ 0 km s$^{-1}$, while dashed (red) contours delineate redshifted gas (integrated over 18 km s$^{-1}$ $\leqslant$ V$_{LSR}$ $\leqslant$ 48 km s$^{-1}$). The dashed rectangle indicate the extent of the mapped region, centered at \textit{$\alpha$} = 18$^h$29$^m$49$^s$.7, \textit{$\delta$} = 01$^o$15$\arcmin$22$\arcsec$.8 (J2000) Filled and dashed controus start at 0.8 K km s$^{-1}$ with an 1 K km s$^{-1}$ increment. HARP-B detector beam of $\sim$14$\arcsec$  is represented as circle in the lower right corner. The dashed rectangle delineates the observed area, and solid rectangles named A, B, C, represent areas that are presented in magnification in Figures \ref{fig:areaA}, \ref{fig:areaB}, and \ref{fig:areaC} respectively. }
\label{fig:field}
\end{figure*}

\begin{figure}
\centering
\resizebox{\hsize}{!}{\includegraphics{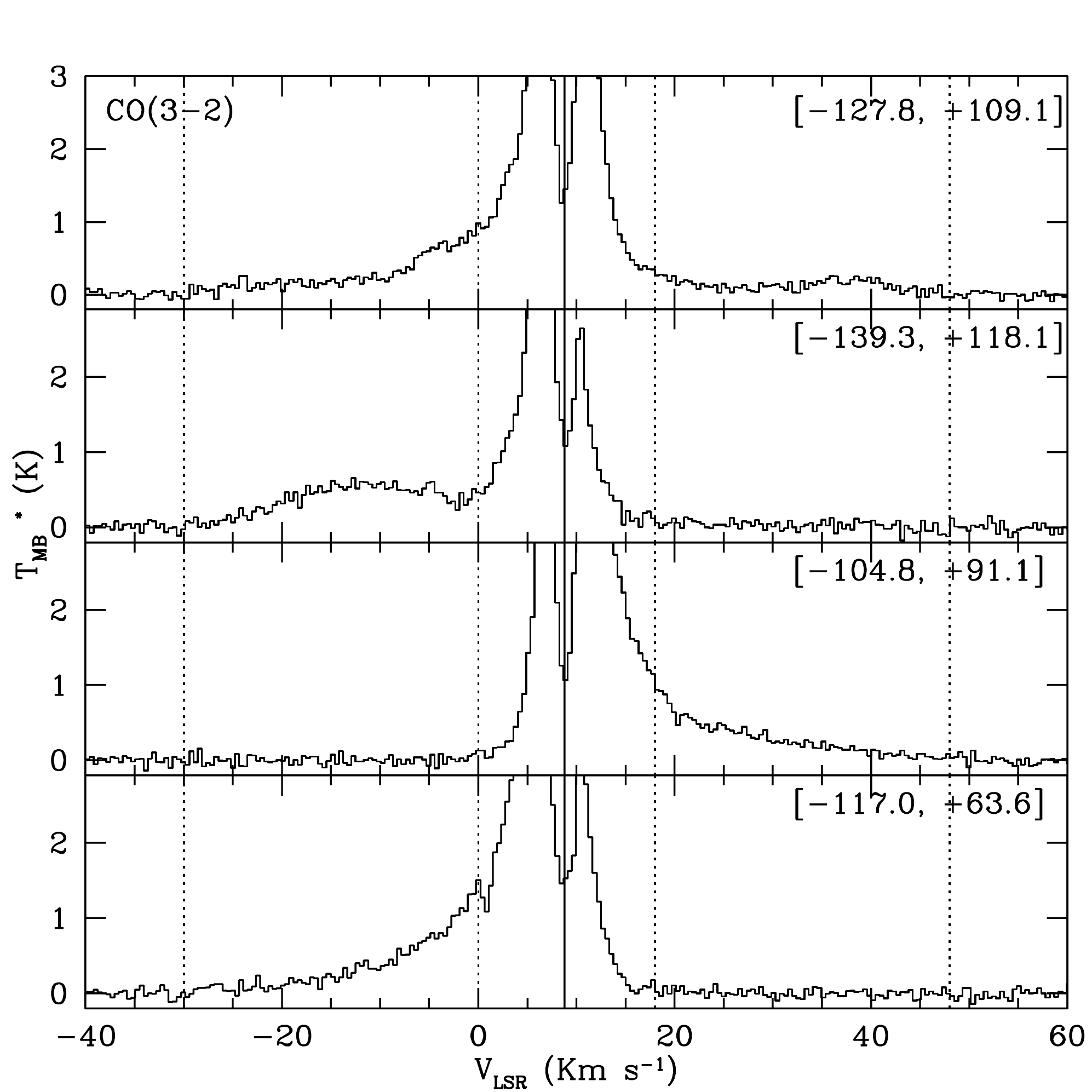}}
\caption {CO \textit{J} = 3$\rightarrow$2 spectra centered at the peaks of the high velocity lobes of area B (see Fig. \ref{fig:field}). Vertical scale is in units of corrected main beam brightness temperature and map offsets are given in the upper-right corner of each spectrum. Dotted vertical lines mark the outer (-30, 48 km s$^{-1}$) and inner (0, 18 km s$^{-1}$) bounds, over which we integrated the high velocity blueshifted and redshifted emission; solid line marks the ambient cloud velocity (V$_{LSR}$ = 8.8 km s$^{-1}$ ). At the lower two panels, high velocity gas forms continuus, fading-off wings, whereas in the upper two panels high velocity gas exhibits a semi-detached distribution with a secondary peak.} 
\label{fig:spectra}
\end{figure}

The global morphology of the high-velocity gas in the investigated region, can be evidenced in Figure \ref{fig:field}, presenting the CO \textit{J} = 3$\rightarrow$2 intensity map integrated over the high velocity blue shifted (-30 km s$^{-1}$ $\leqslant$ V$_{LSR}$ $\leqslant$ 0 km s$^{-1}$) and red shifted (18 km s$^{-1}$ $\leqslant$ V$_{LSR}$ $\leqslant$ 48 km s$^{-1}$)  wings,  superimposed over a (grayscale) Spitzer-MIPS 24 $\mu$m image.  The 24 $\mu$m band was selected for being sensitive in tracing continuum emission from protostellar envelopes surrounding deeply embedded sources, and providing superior spatial resolution than the other, longer-wavelength MIPS bands.  In Figure \ref{fig:field}, the dashed rectangle delineates the extent of the mapped area, and solid line rectangles A, B and C indicate areas that are discussed in detail in \S\S \ref{areaA} - \ref{areaC}.  CO \textit{J} = 3$\rightarrow$2 representative spectra sampling the map of Fig. \ref{fig:field} at different offsets are displayed in Figure \ref{fig:spectra};  dashed vertical lines mark the velocity intervals over which the intensity is integrated, and the solid line indicates the cloud systemic velocity ($\sim$ 9 km s$^{-1}$), coinciding with the self-absorption profile. Low velocity cuts at $\pm$9 km s$^{-1}$ from the body of the line ensures that the ambient contribution is excluded. High velocity cuts are imposed by the rms noise level of the maps; the higher sensitivity obtained with these observations \citep[$\sim$ 4 times higher than the one reached by ][]{Davis} with the same the same telescope) allow to extend the high velocity cutoffs to $\sim \pm$ 40 km s$^{-1}$ from the V$_{LSR}$. 

In Fig. \ref{fig:spectra}, two types of high velocity wing morphologies are distinguishable:  in the lower two panels, high velocity gas forms continuous, fading-off wings as a function of velocity from V$_{LSR}$;  in the upper two panels high velocity gas exhibits a semi-detached distribution, with a secondary peak tracing clumps of high velocity blue and red-shifted gas. The latter wing morphology is indicative of high velocity "bullets" of gas, commonly observed in molecular jets from Class 0 protostars \citep[e.g.][]{Bachiller2} 

The large scale spatial distribution of the poorly evolved Class0/I young stellar objects in Serpens, identified on the basis of their SED \citep{Hurt1, Larsson, Winston}, infrared excess and spectral indices \citep{Kaas1, Kaas2, Enoch2, Winston}, or detection of luminous millimeter and sub-millimeter continuum sources \citep{Casali, Testi1, Davis, Williams}, displays concentrations into two clumps located at the NW (areas A and B) and SE (area C), separated by $\sim$ 200$\arcsec$ \citep{Casali}, as traced by the 24$\mu$m MIPS background image in Figure \ref{fig:field}. Not surprisingly, most of the outflow activity observed in same figure is tightly constrained to the same NW and SE clumps, as powerful molecular outflows are known to be driven by very young stellar objects \citep{Richer}. Twelve out of fourteen sources classified as bona-fide Class0/I sources \citep{Winston} lie within the extent of the  CO \textit{J} = 3$\rightarrow$2 HARP-B map. 

The distribution of more evolved Class II sources in contrast, shows no signs of  clustering  but are found to be more dispersed \citep{Kaas1}. It has been suggested that the difference in the spatial distribution of YSOs at different evolutionary stages reflects different star forming episodes in Serpens \citep{Casali, Hurt1, Kaas2}. 

In \S\S \ref{areaA} - \ref{areaC} we examine the particular characteristics for each one of the defined areas A-C and attempt to attribute the observed outflows to individual sources. From this perspective, isolated outflows that are not comprised within the examined areas could not be connected to any of the protostellar sources in the mapped area. High velocity CO \textit{J} = 3$\rightarrow$2 intensity maps are examined in conjuction with Spitzer IRAC band-2 (4.5 $\mu$m) and MIPS band-1 (24 $\mu$m) images from \citet{Winston}. Within the  IRAC band-2 fall a number of high-excitation energy pure rotational and ro-vibrational H$_2$ lines, which trace warm, shocked gas from the underlying protostellar jets; such emission assists in the coupling between outflows and candidate driving sources. Possible connections are also derived  comparing with data from the very rich literature on Serpens, that covers observations of near-IR jets to high density sub-millimeter and millimeter molecular outflow tracers. From this analysis, we identify 11 sources that most likely are responsible for the bulk of molecular outflow activity observed in the Serpens Core. About half of sources appear to drive bipolar outflows and from these, four exhibit  H$_2$ jet-like emission associated with these outflows.  The projected position angles of outflows, estimated from the peaks of outflow emission with respect to the corresponding driving sources,  tend to have, within uncertainties mainly due to the limited resolution of the maps, an orientation in the NW-SE direction (see Table \ref{tab:1} and Figures \ref{fig:areaA}, \ref{fig:areaB}, and \ref{fig:areaC}). This alignemnt that has been attributed to influence of the helical component of the large scale magnetic field \citep{Gomez}, however recent polarimetry observations \citep{Sugitani} indicate a hourglass shaped field centered between the NW and SE clumps, which in many cases is aligned perpendicularly to the direction of the observed outflows. Therefore no secure conclusions can be drawn from the present data on the role of the local magnetic field and the outflow alignment.

For the positional determination and characterization of the evolutionary state of the YSO's in the present study, we have employed the analytical catalogues of \citet{Winston}. To avoid confusion, when existing we adopt the nomenclature of SMM sources listed in  \citet{Casali}, whereas for the rest we use the \textit{Spitzer} ID from the \citet{Winston} catalogue with the letter \textit{W} as a prefix


\subsection{Area A}
\label{areaA}

\begin{figure}
\resizebox{\hsize}{!}{\includegraphics{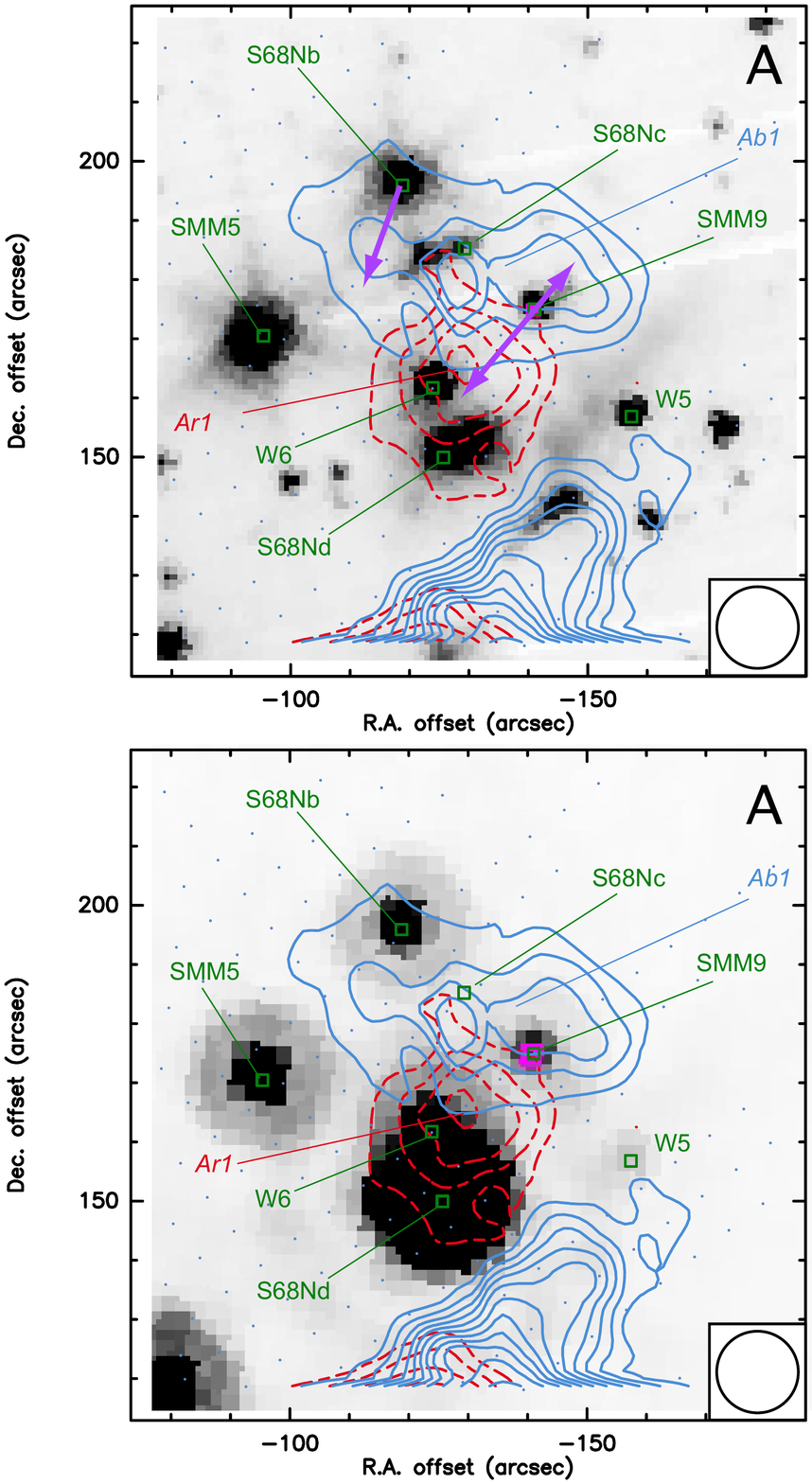}}
\caption{Magnification of the high velocity gas of Area A from Figure \ref{fig:field} superimposed on an  IRAC band 2 (upper panel) and MIPS band 1 (lower panel) images (in grayscale). In the upper panel,  arrows indicate outflow axis and possible associations with protostellar sources. Contour levels are as in Figure \ref{fig:field}}
\label{fig:areaA}
\end{figure}

\begin{figure*}
\centering
\resizebox{14cm}{!}{\includegraphics{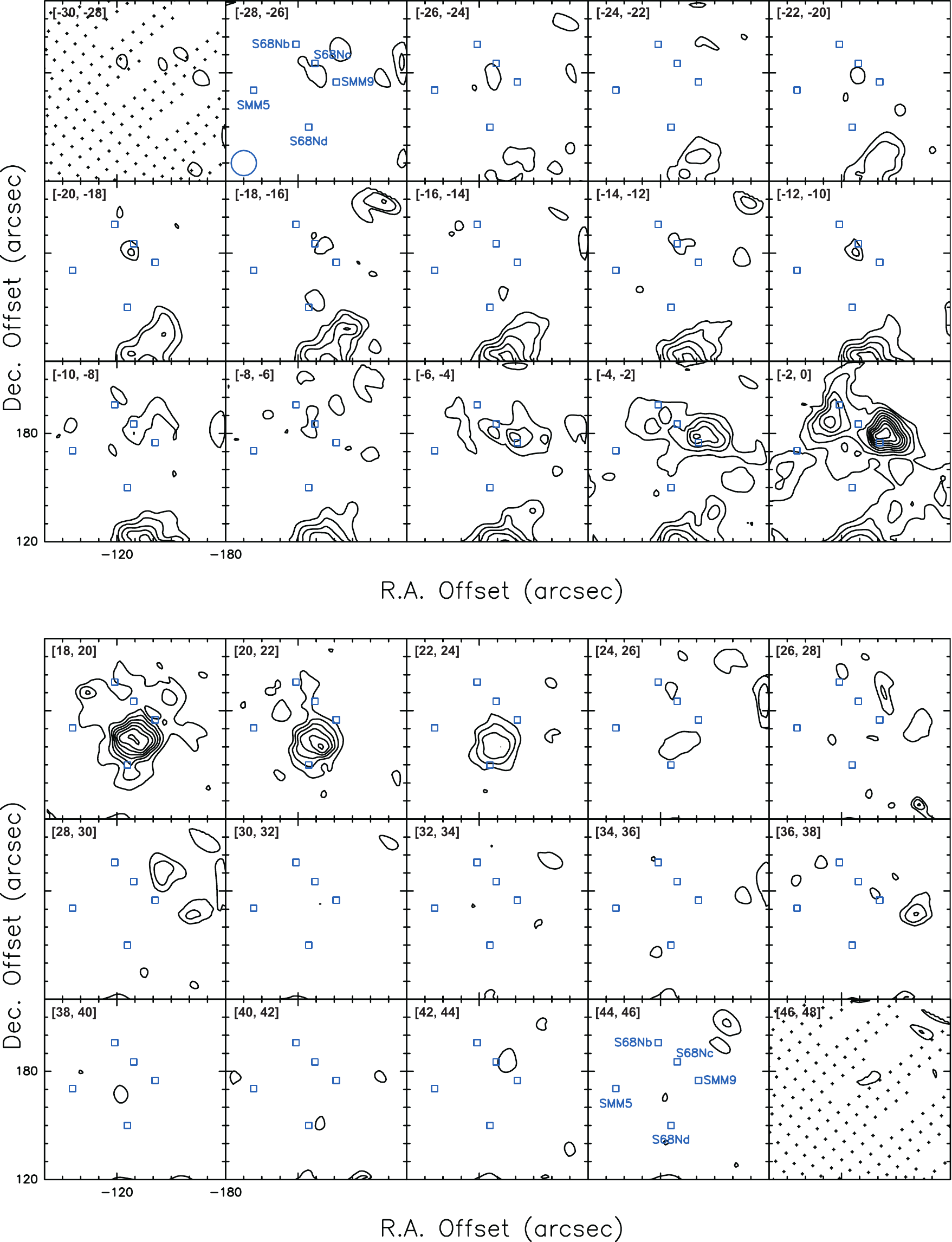}}
\caption{Velocity channel map of area A (see Fig. \ref{fig:areaA}). Each panel displays emission integrated over a 2 km s$^{-1}$ interval, and contour levels start from 0.25 K km s$^{-1}$ with a 0.25 K km s$^{-1}$ increment. Crosses in the first and last panels display the sampling; the position of sources likely to drive outflows are displayed with empty squares}
\label{fig:vcmA}
\end{figure*}

\begin{figure}
\centering
\resizebox{\hsize}{!}{\includegraphics{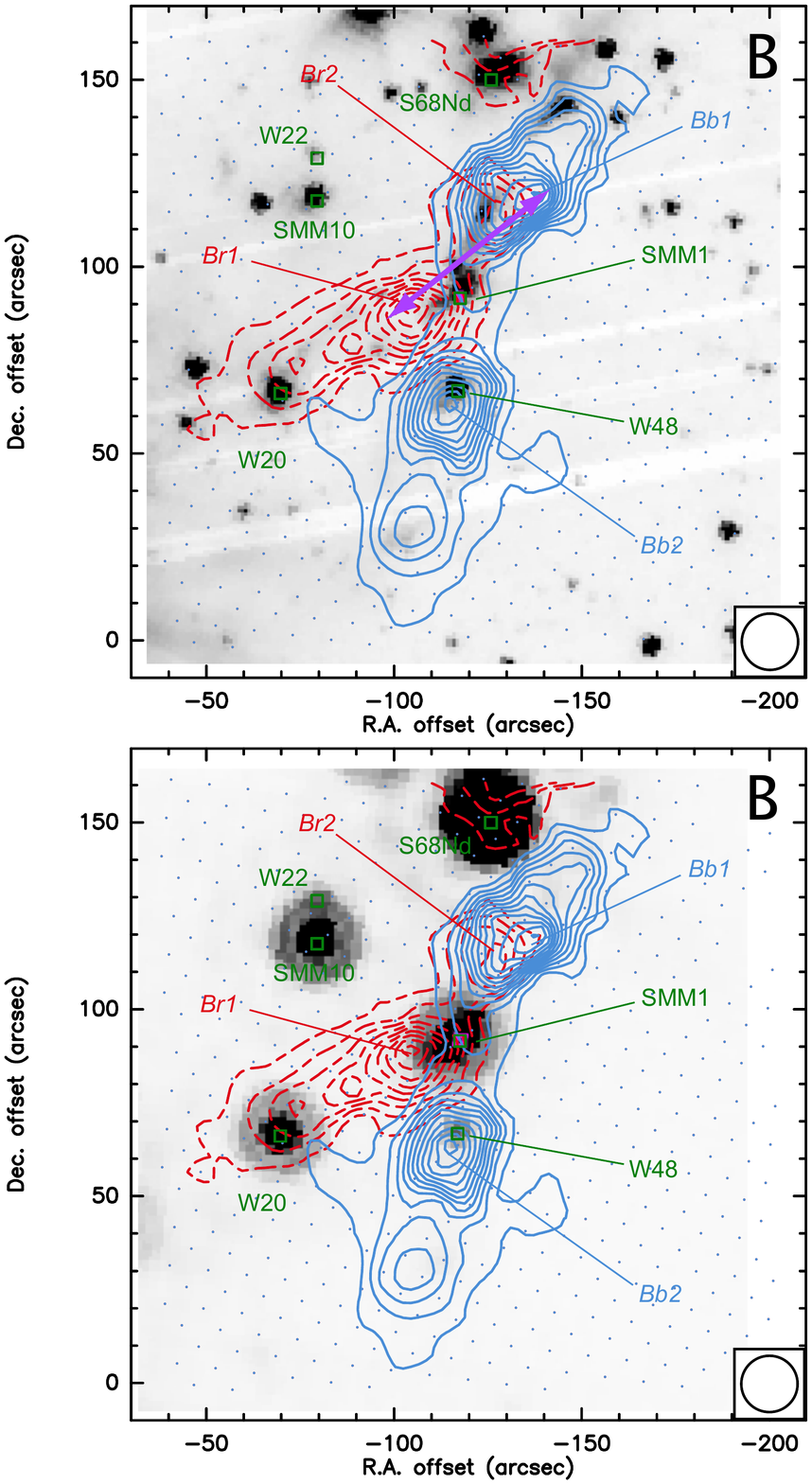}}
\caption{Magnification of the high velocity gas of Area B from Figure \ref{fig:field} superimposed on an  IRAC band 2 (upper panel) and MIPS band 1 (lower panel) images (in grayscale). Contour levels are as in Figure \ref{fig:field}}
\label{fig:areaB}
\end{figure}

Figure \ref{fig:areaA} presents a magnification of the outflow activity within Area A (Fig. \ref{fig:field}) superimposed over an IRAC band 2 (upper) and MIPS band 1 (lower) grayscale images while Figure \ref{fig:vcmA} shows individual velocity channel maps, that have assisted in the identification of the
outflowing gas with respect to the ambient material, and in the association of blue- and red-shifted
outflow lobe pairs emitted from the same source.  In both panels of Fig. \ref{fig:areaA} the positions of known Class0/I sources from the catalogue of \citet{Winston} are labeled along with prominent high velocity outflow lobes; the extended blue-shifted emission at the south is discussed in the next section. Being located close to the border of the CO map, area A is not fully sampled; contour levels employed here are the same as in Fig. \ref{fig:field} for the sake of consistency, however outflows are better delineated in the velocity channel maps close to the low velocity cutoffs of Fig. \ref{fig:vcmA}, where contour levels start at 0.25  K km s$^{-1}$ and are spaced by the same amount. 

Area A covers the S68N region at the northwest of the Serpens Core;  submillimeter single dish \citep{Davis} and interferometric observations \citep{Testi1, Williams} have revealed the existence of four cores, named S68Na through d, with S68Na most likely coinciding with the continuum source SMM9 \citep{White, Wolf-Chase}. To the east, the continuum source SMM5 \citep{Casali} is possibly associated with the bright near-IR cometary nebula EC53 \citep{Eiroa3, Hodapp}; however, while the head of the cometary nebula lies close to the sub-mm source, the latter is not detected in the near-IR.

The blue and red shifted lobes Ab1 and Ar1 in Figure \ref{fig:areaA} most likely represent a bipolar outflow from S68Na/SMM9 at a projected position angle (P.A.) of $\sim$ 150$^o$.  Although Ab1 being at the far edge of the map is not well outlined, the jet like emission in the IRAC image adjacent to the driving source and pointing to the NE possibly indicates the direction of the underlying jet, which in this case coincides with the orientation of the associated outflow.  Maps of CS \textit{J} = 2$\rightarrow$1 \citep{Wolf-Chase, Testi2, Williams}, methanol  \citep{Testi2, Garay} and SiO \citep{Garay} outline a similar bipolar outflow connected to SMM9 at a position angle of $\sim$ 135$^o$-145$^o$, and therefore are in support to such correlation.

The extension of lobe Ab1 to the east lying roughly between S68Nb and SMM5 is harder to associate. The underlying jet-like emission in the IRAC image very close to S68Nc pointing to the SE is resolved as a series of knots in near-IR  H$_2$ 1-0 () \textit{S}(1) (2.12 $\micron$) imaging \citep{Hodapp}; proper motions  show that these knots are moving towards SMM5 \citep{Hodapp}, therefore favoring S68Nc as their progenitor. Still, it is doubtful if there exists an embedded protostar within the S68Nc clump, as it is scarcely detected in the far-IR and sub-mm continuum \citep{Winston, Davis}, and only appears as an elongated condensation at 3 mm \citep{Williams}. In the [-2, 0] velocity channel map of Fig. \ref{fig:vcmA}, Ab1 displays an orientation towards the south, implying a possible connection with S68Nb; as other sources do not provide convincing evidence,  we  tentatively associate the extension of lobe Ab1 to the source S68Nb.   

In the [-2,0] and [18, 20] channels of Figure \ref{fig:vcmA}, two weak (blue and red-shifted) lobes pointing to the SE and NW from SMM5 and possibly associated to this source are recorded. Due to the contour level selection, these lobes are not visible in  Fig. \ref{fig:areaA}, however we consider them as real and name them Ab2 and Ar2 (see table \ref{tab:2}). Similar weak red-shifted emission with roughly the same orientation is only recorded in the \textit{J} = 1$\rightarrow$ CO map of \citet{Narayanan}.  

\subsection{Area B}
\label{areaB}

Area B is located to the south of Area A and displays the outflow activity below the S68N clump. Although Areas A and B together confine the Serpens NW region, due to its unique characteristics this segment is presented here stand-alone.  As in the case of Area A, in Figure \ref{fig:areaB} we present high velocity \textit{J} = 3$\rightarrow$2 CO outflows superimposed on the IRAC band 2 and MIPS band 1 images (upper and lower panels respectively). Again, deeply embedded Class0/I sources are tagged along with the strongest outflow lobes. 

This area comprises SMM1/FIRS1 and SMM10, the former being most prominent source in Serpens \citep{Casali, Hurt1, Davis}, that displays a series of mass ejection phenomena. It has been associated with a radio jet and OH maser emission \citep{Rodriguez, Curiel1} while strong outflows have been traced in a number of molecular lines. 

\begin{figure*}
\centering
\resizebox{14cm}{!}{\includegraphics{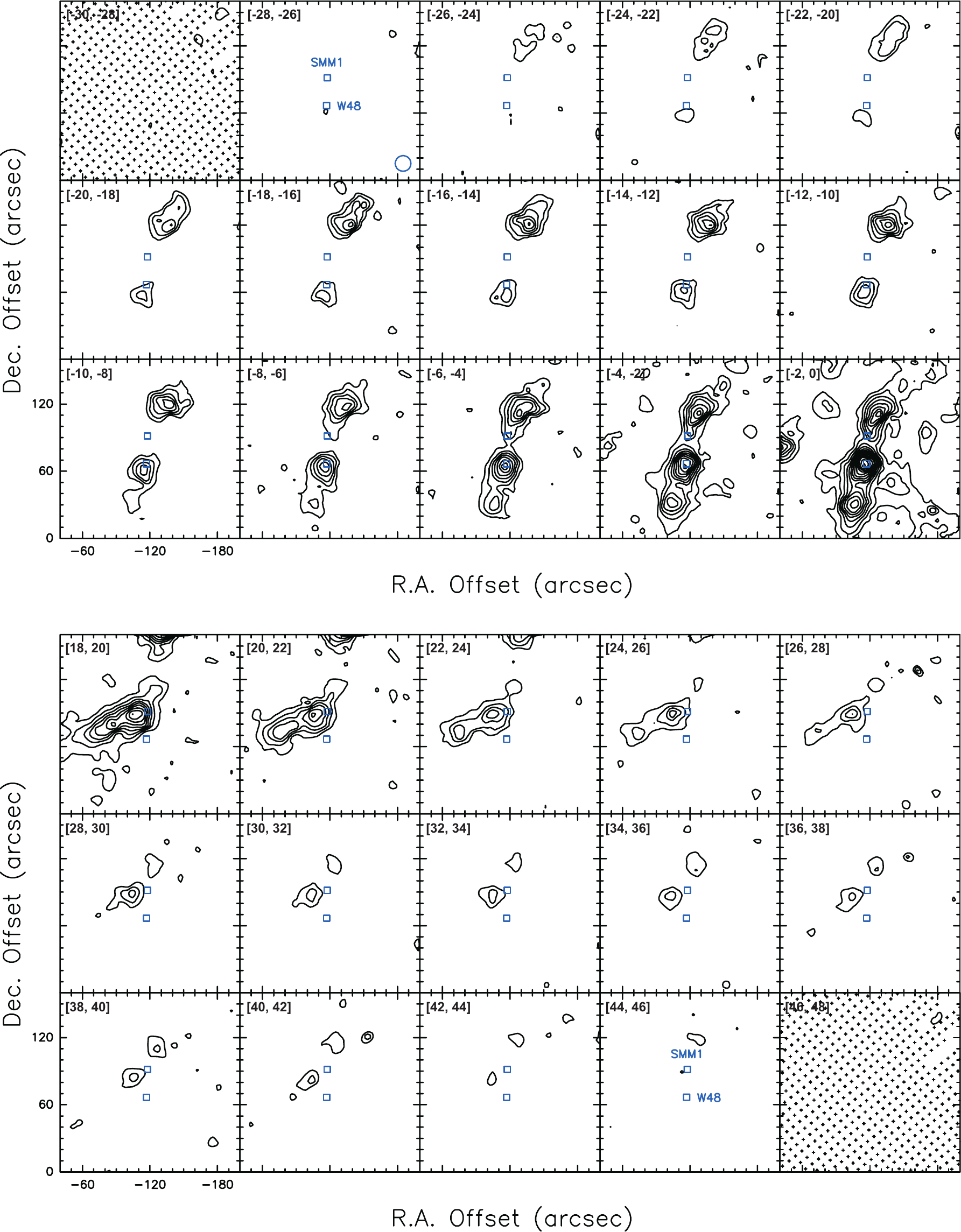}}
\caption{As in Fig. \ref{fig:vcmA}, for the area B (see Fig. \ref{fig:areaB})}
\label{fig:vcmB}
\end{figure*}

High velocity CO emission as evidenced in Figure \ref{fig:areaB} is rather puzzling; there are two blue-shifted (Bb1,2) and two red-shifted lobes (Br1,2) that all seem to converge towards SMM1. Both Bb1 and Br2 point out towards the NW and overlap; the line wing morphology at the peaks of these lobes, presented in the upper two spectra of Fig. \ref{fig:spectra} indicates the existence of high velocity "bullets", which in velocity channel maps of Figure \ref{fig:vcmB} can be traced up to the [-24,-22] and [+38,+40] velocity bins, symmetrically to the nominal cloud velocity. The spatial coincidence along with the common wing morphology suggest that Bb1 and Br2 represent the same high velocity gas projected at different Doppler-shifts on the plane of the sky.

 \begin{table}[!t]
\caption{Outflow position angles} 
\label{tab:1} 
\centering 
\begin{tabular}{l  c} 
\hline\hline 
Source		& 	Outflow P.A. ($^o$)  \\	
\hline
SMM1$^a$ & 130 $\pm$ 5 \\
SMM3$^a$ &  155 $\pm$ 8 \\
SMM4$^a$ & 170 $\pm$ 7 \\
SMM5$^a$ & 135 $\pm$ 10 \\
SMM6$^b$ & 35 $\pm$ 8\\
SMM8$^c$ & $\ldots$ \\
SMM9$^a$ & 150 $\pm$ 10 \\
S68Nb$^b$ & 160 $\pm$ 10\\
W10$^b$   &  130 $\pm$ 8 \\
W45$^a$ & 70 $\pm$ 15 \\
W48$^c$ & $\ldots$ \\ 
\hline

~$^a$ driving bipolar outflow\\
~$^b$ driving unipolar outflow\\
~$^c$ position angle uncetain\\

\end{tabular}
\end{table}

Mid-IR IRAC emission in Figure \ref{fig:areaB}, shows arc-shaped features, extending to the NW and SE outlining the rim of the CO lobes Bb1 and Br1. The same morphology is also traced in  Near-IR 1-0 \textit{S}(1) H$_2$ imaging \citep{Hodapp} and possibly in CS \citet{Williams}; we note that unlike the outflow from SMM9, the CS features corresponding to Bb1 and Br1 have opposite Doppler-shift than the CO lobes and are much weaker. The base of the NW and SE cavities very close to SMM1 also appear in high density molecular tracers such as HCO$^+$ and HCN \citep{Hogerheijde} and the (1,1) transition of ammonia \citep{Torrelles}.

All the existing evidence leads to the association of Bb1 and Br2 pointing to the NW and Br1 to the SE with SMM1 at a single bipolar outflow configuration; however, more complex configurations involving a binary source \citep[as suggested by][]{White} cannot be excluded. The position angle of the outflow is $\sim$130$^o$, coinciding with the position angle of $\sim$ 125$^o$ of the underlying radio jet  \citep{Rodriguez, Curiel1}, a fact evidencing the possible connection between the two phenomena. To the south, the alignment of the peak of lobe Bb2 with the flat spectrum source W48 \citep{Winston} may be interpreted as if Bb2 is a face-on unipolar outflow with a non-detected receding, red-shifted lobe pointing away from the observer.

Very little evidence of outflows exist in the vicinity of SMM10, traced only in the  [-2, 0] channel of Figure \ref{fig:vcmB}. The evolutionary stage of this protostellar source is uncertain; \citet{Eiroa} has classified it as a Class 0 source, but the significant emission recorded shortward 24 $\micron$ \citep{Winston} and the lack of strong outflows are indicative of a more advanced protostellar stage. 

\subsection{Area C}
\label{areaC}

Area C comprises the SE clump of the Serpens Core, a highly populated region where protostellar sources in various evolutionary stages are packed together with bright near-IR nebulosities. Millimeter and sub-millimeter maps \citep{Casali, Testi1, Davis} reveal four significant condensations, named after \citet{Casali} as SMM2,3,4 and 6. In the near-IR, the bright Serpens Reflection Nebula (SRN) is associated with SVS20 and SVS2, the two brightest near-IR sources in the Serpens Core complex \citep{Eiroa3}.

\begin{figure}
\resizebox{\hsize}{!}{\includegraphics{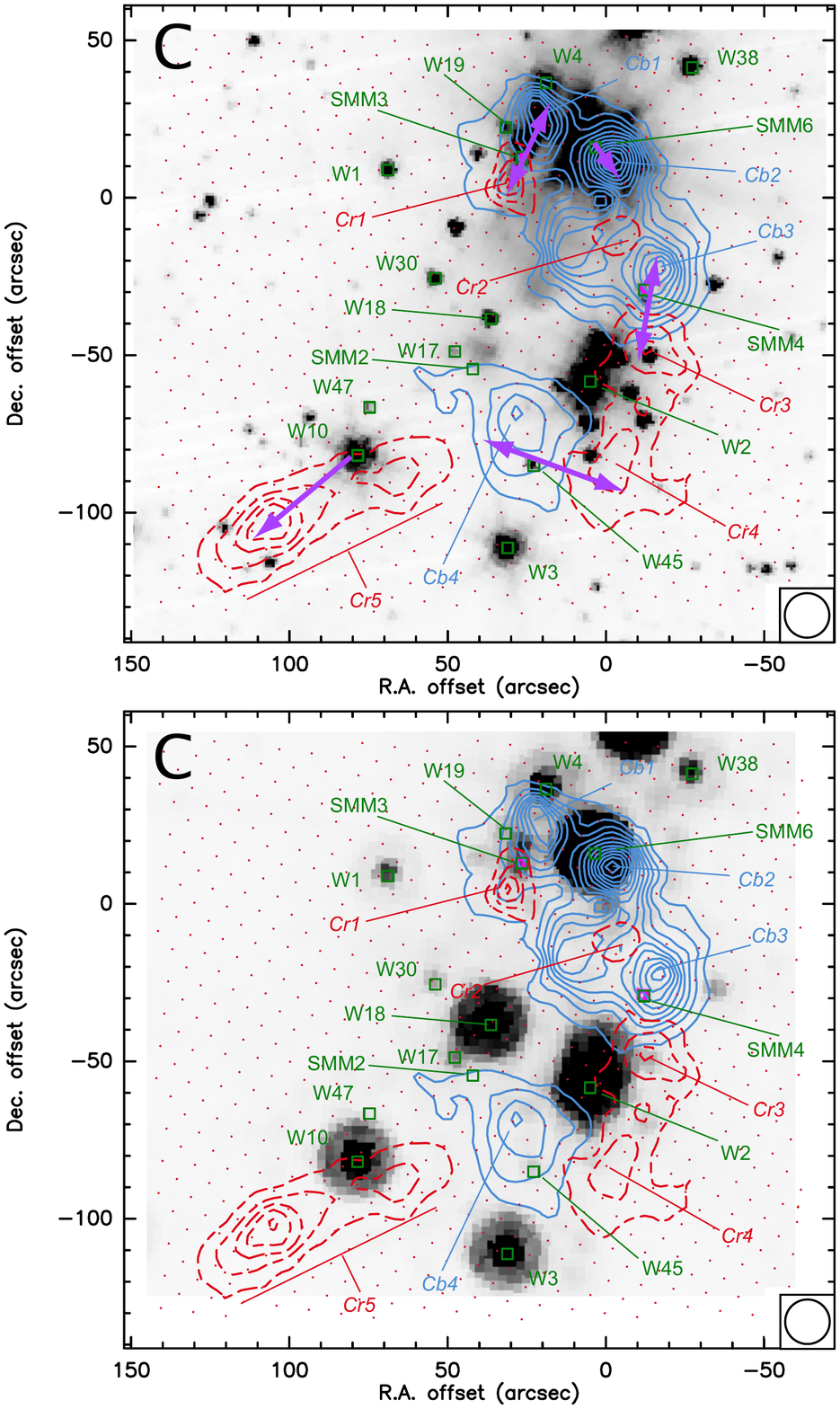}}
\caption{Magnification of the high velocity gas of Area C from Figure \ref{fig:field} superimposed on an  IRAC band 2 (upper panel) and MIPS band 1 (lower panel) images (in grayscale). Contour levels are as in Figure \ref{fig:field}}
\label{fig:areaC}
\end{figure}

\begin{figure*}
\centering
\resizebox{14cm}{!}{\includegraphics{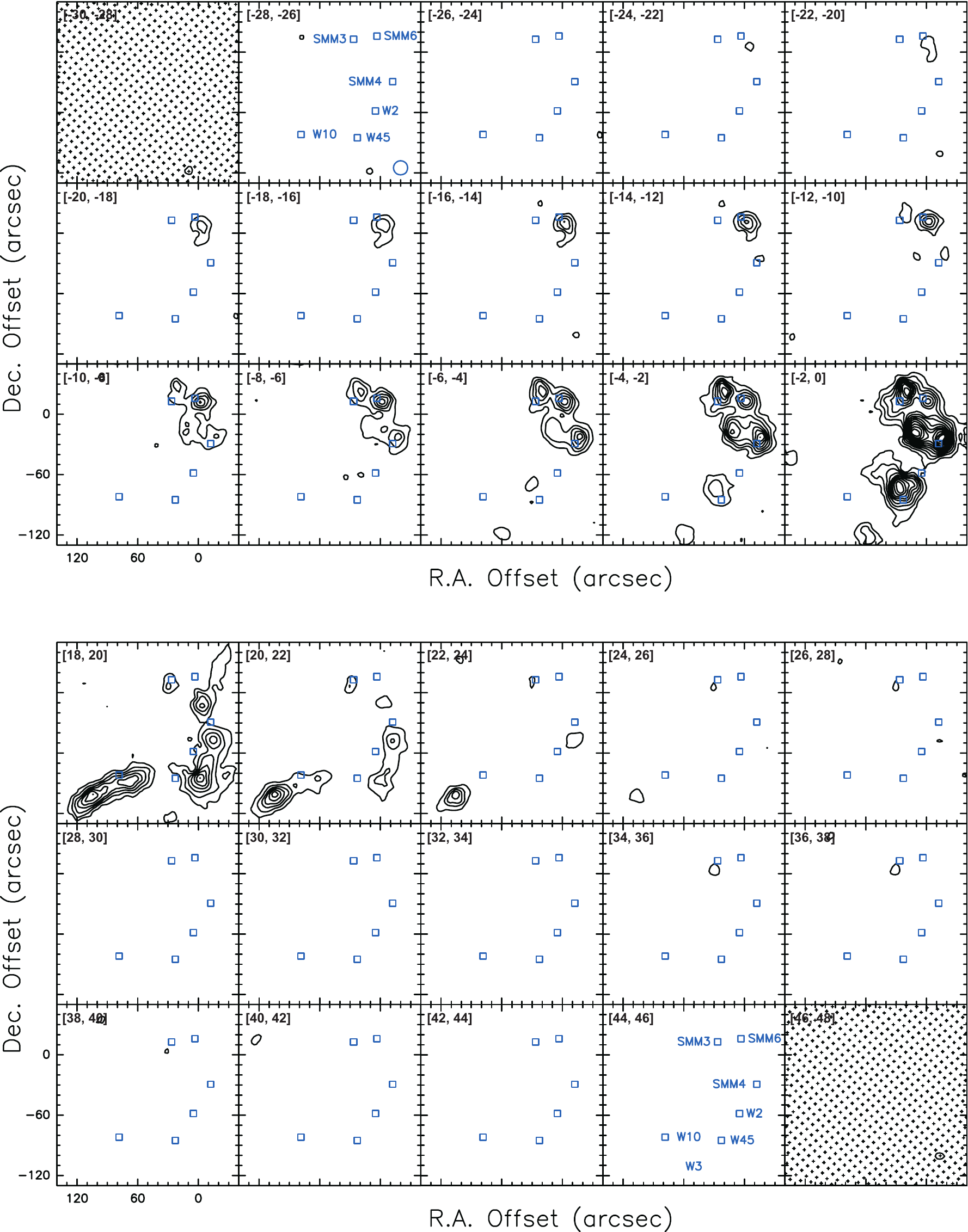}}
\caption{As in Fig. \ref{fig:vcmA}, for the area C (see Fig. \ref{fig:areaC})}
\label{fig:vcmC}
\end{figure*}

As in the previous areas, the high velocity CO \textit{J} = 3$\rightarrow$2 outflows are displayed over the IRAC band 2 (upper) and MIPS band 1 (lower) panels in Figure \ref{fig:areaC}. The positions of the SMM sources along with other Class0/I and flat spectrum sources from \citet{Winston} are marked in both panels. A high concentration of possible Class II sources and field stars is observed towards the SW in the IRAC image, while significant emission from the reflection nebula is apparent towards the north and center of Area C, With the exception of SMM6, all other SMM sources do not appear bright on the  24 $\micron$ MIPS band.  Outflow contours around SMM3,4 and 6 are clustered, but display a number of peaks which are tagged in both panels of Figure \ref{fig:areaC}. The velocity range for the majority of the outflows in this area is lower than observed for the NW clump (areas A and B); significant blueshifted and redshifted emission is recorded up to [-10, -8] and [26, 28] km s$^{-1}$ velocity bins respectively, with an exception of the blueshifted lobe \textit{Cb2} which is observed up to the [-26, -24] km s$^{-1}$ channel (Fig. \ref{fig:vcmC}).

With the exception of SMM2, all the SMM sources in the region have associated CS and H$_2$CO emission counterparts \citep{Mangum}. However, it is not yet clear whether SMM2 harbors a protostellar source. Despite being observed as a condensation in the sub-millimeter \citep{Davis}, it is scarcely or not at all traced in millimeter continuum observations \citep{Testi1, Hogerheijde}.In addition, the two nearby mid-IR Class0/I sources W17 and W18 present in the 24 $\micron$ MIPS band \citep{Winston} are offset from the 3-mm position of SMM2 \citep{Testi1}, and therefore not related to it. The lobe Cb4, located 20$\arcsec$ to SW is most likely not related to SMM2 but with some of the more evolved sources to the SE, possibly in a bipolar scheme with Cr4. Therefore, the absence of any apparent outflow emission in the vicinity is in favor of the view that SMM2 is a warm cloud condensation without an embedded source.

SMM3 is a mm/sub-mm continuum source located to the north of Area C; in the IRAC image, there is strong jet-like emission extending to the NW from the source, at a position angle of 155$^o$ traced also in the near-IR H$_2$ 1-0 S(1) line \citep{Herbst, Eiroa2}. Most likely, the emissions observed to the same direction in  HCO$^+$, HCN and SiO \citep{Hogerheijde} and CS \citep{Testi2} trace the same outflow activity.  Lobe Cb1 follows the direction of the jet-like emission and peaks $\sim$ 10$\arcsec$ to the NW of SMM3, while Cr1 is collinear, pointing to the opposite SE direction and peaks roughly at the same projected distance from the source. Even though a number of other nearby Class0/I sources (e.g. W19, W4) are present nearby, the positioning of SMM3 with respect to Cb1 and Cr1  along with the outflow detected in other molecular tracers are in favor of their association.

SMM4 is located to the west of Figure \ref{fig:areaC}, between the two dominant diffuse emission regions in the IRAC image. In the same image, two small nebulosities appear very close to the source in a North-South direction. The same morphology is observed in the near-IR \citep{Eiroa}, while outflows with the same direction are traced in CS, CH$_3$OH \citep{Testi2}, HCO$^+$ and HCN \citet{Hogerheijde}, as well as in lower \textit{J} CO maps \citep{Davis, Narayanan}   This structure becomes evident in the \textit{J} = 3$\rightarrow$2 map of \citet{Narayanan} where a bipolar outflow is clearly shaped. In Figure \ref{fig:areaC}, lobes Cb3 and Cr3 display similar morphological characteristics, and most likely are  associated with SMM4.

To the NW of Area C, the bright IRAC nebulosity is at the location of the near-IR binary SVS20/EC90, the brightest near-IR source in Serpens \citet{Eiroa3} coinciding with the position of mm/sub-mm source SMM6.  This is a binary system is surrounded by a near-IR ring or nebulosity, with two spiral arms seen in opposite directions \citep{Herbst, Eiroa2}, that have been interpreted as a circumbinary disk \citep{Eiroa2} or evacuated bipolar cavities \citep{Huard}. Binary components have a north-south distribution separated by only 1.5$\arcsec$ and both present a flat spectrum SED \citep{Haisch}. It has been suggested that the southern component has strong Br$\gamma$ and CO bands in emission indicating the presence of an accretion disk that most likely corresponds to SMM6 \citep{Eiroa}.  The most prominent southern arm observed in the near-IR H$_2$ emission \citep{Huard, Herbst, Eiroa2} is in the direction of the blue-shifted lobe Cb2 in Figure \ref{fig:areaC}, and we tentatively attribute it to SMM6. Most likely, the red-shifted emission is obscured by the nebulosity, as A$_V$ values ranging between 9 mag \citep{Cambresy} and more than 20 mag \citep{Huard} have been measured for this region.

At the south of  Area C, there is a wealth of IRAC sources, the majority of them classified as Class II protostars or field stars. Class 0/I and flat spectrum sources from the \citet{Winston} catalogue are labeled on Figure \ref{fig:areaC}, around lobes Cb4, Cr4 and Cr5. These lobes are present in lower \textit{J} CO observations but are mostly prominent in the high velocity map of \citet{Davis}. The protostellar source W45 lying approximately between lobes Cb4 and Cr4 is a flat spectrum source \citep{Winston} that could potentially drive the observed bipolar structure. Even though it is not the least evolved nor the best positioned source with respect to the outflows, it is the only one optimally satisfying both conditions. Similarly, the red-shifted lobe Cr5 is located just south of W10, classified as Class0/I in \citet{Winston} catalogue which is possibly correlated to the bright continuum source PS2 detected by \citet{Hurt1} and also present in the 3-mm map \citet{Testi1} . In the absence of any blue-shifted emission in the vicinity, we tentatively attribute Cr5 to W10 as a unipolar outflow. 

Strong  isolated red-shifted emission is evident in Figure \ref{fig:field} to the north but outside of Area C in the vicinity of mm/submm source SMM8. This emission corresponds to the NEr lobe of the high velocity CO map of \citet{Davis}, and as in the case of W10, we tentatively correlate this emission to SMM8 as a unipolar outflow. Finally, no possible driving sources could be found for the remaining blue and red-shifted lobes to the upper and lower left (east) of Area C in Figure \ref{fig:field}. We have examined thoroughly the literature for a possible source between lobe Cr5 (Area C) and the intense blue-shifted emission outside the lower left corner of Area C which corresponds to lobe SEb in the high velocity maps of \citet{Davis} that could justify a possible bipolar scheme, but no appropriate candidate sources could be found.

\begin{table*}
\caption{Outflow properties, for the blue (integrated over -30 km s$^{-1}$ $\leqslant$ V$_{LSR}$ $\leqslant$ 0 km s$^{-1}$) and red-shifted (integrated over 18 km s$^{-1}$ $\leqslant$ V$_{LSR}$ $\leqslant$ 48 km s$^{-1}$ ) gas; lower and upper limits correspond to calculations based on the two adopted values for $\tau$=0,1}. 
\label{tab:2} 
\begin{tabular}{l c c c c c c c c c} 
\hline\hline 
Outflow  &  Source & N$_{tot}$   &  M$_{tot}$   &  P$_{tot}$   &       E$_{K}$ &   t$_{dyn}$ & F$_{CO}$ & L$_{mech}$\\
&   &  (10$^{16}$cm$^{-2}$)    &   (10$^{-3}$  $M_{\sun}$)   &  (10$^{-2}$ $M_{\sun}$ km s$^{-1}$)   &   (10$^{42}$ erg)      & (10$^{3}$ years) &  (10$^{-5}$ $M_{\sun}$ km s$^{-1}$ yr$^{-1}$) & (10$^{-2}$ L$_{\sun}$)\\

\hline
\multicolumn{8}{c}{Area A}\\
\hline
Ab1 	&	SMM9	& 	1.89 - 2.97				 & 	0.49 - 0.78 	& 		1.03 - 1.63		 & 	1.25 - 1.98  	  	&	0.7	&		1.47 - 2.33    & 1.42 - 2.26   \\
Ar1 	&	SMM9	& 	1.87 - 2.95				 & 	0.48 - 0.77 	& 		1.10 - 1.72		 & 	0.73 - 1.16  	  	&	0.9	&		1.22 - 1.72    &  0.64 - 1.03  \\
Ab2	& 	SMM5	& 	1.53 - 2.41				 &	0.4  - 0.63    	&		0.3 - 0.49		 	 &	0.24 - 0.4      	  	&	2.9	&		0.11 - 0.17    &  0.07 - 0.11  \\	
Ar2	& 	SMM5	& 	1.07 - 1.7				 	 &	0.3  -  0.45   	&		0.23 - 0.37		 & 	0.2  - 0.32     	  	&	1.5	&		0.15 - 0.24    &  0.11 - 0.17  \\	
Ab3	& 	S68Nb	& 	2.49 - 3.94				 &	0.6  -  1.03   	&		0.5 - 0.8		 	 & 	0.41 - 0.65     	  	&	1.8	&		0.27 - 0.44    &  0.18 - 0.29  \\
\hline
\multicolumn{8}{c}{Area B}\\
\hline
Bb1 	&	SMM1	& 	10.44 - 16.52	 		 	 & 	2.73 - 4.33			& 		5.64 - 8.92 	 & 	10.78 -  17.04	  	&	2.6	& 		2.57 - 3.43   & 3.31 - 5.24	\\
Br1 	&	SMM1	& 	8.62 - 13.63		 		 &	2.26 - 3.57			& 		4.22 - 6.67 	 & 	6.23 -  9.85     	  	&	2.9	&		1.45 - 2.30   & 1.71 - 2.71    \\
Br2	 &	SMM1	& 	1.86 - 2.94		 		 & 	0.49 - 0.77			& 		1.23 - 1.95 	 & 	2.42 - 3.82     	 	&	1.5	&		0.81 - 1.30  &  1.29 - 2.03  \\
Bb2$^a$  & 	W48	& 	10.02 - 15.84		 	 	 & 	2.62 - 4.15			& 		4.36 - 6.89 	 & 	5.97 - 9.44  	 	&  	$\ldots$&		$\ldots$        	    &    $\ldots$ \\
Bb2$^b$ &	W48	& 	7.47 - 11.81				 &	1.96 - 3.09			& 		3.28 - 5.19 	 & 	5.05 - 7.98         	&	$\ldots$&		$\ldots$    	    &    $\ldots$   \\
\hline
\multicolumn{8}{c}{Area C}\\
\hline
Cb1	&	SMM3	& 	3.27 - 5.17				 & 	0.86 - 1.35			& 		1.29 - 2.04		 & 	1.64 - 2.59  	  	&	1.1	&		1.17 - 1.85 	& 	1.17 - 1.88       \\
Cr1 	&	SMM3	& 	0.68 - 1.08				 & 	0.18 - 0.28			& 		0.42 - 0.66		 & 	0.60 - 0.94  	  	&	0.9	&		0.47 - 0.73     	& 	0.53 - 0.83  \\
Cb3 	&	SMM4	& 	7.45 - 11.79				 & 	1.95 - 3.09			& 		2.96 - 4.68		 & 	3.78 - 5.97  	  	&	0.7	&		4.24 - 6.7       	&	4.32 - 6.82 \\
Cr3 	&	SMM4	& 	2.04 -	3.22	  			 & 	0.53 - 0.84			&		1.07 - 1.69		 & 	0.82 - 1.29     	  	&	1.5	&		0.70 - 0.79       	&	0.43 - 0.62 \\
Cb2 	&	SMM6	& 	4.52 - 7.15				 & 	1.18 - 1.87			& 		2.21 - 3.49		 & 	4.02 - 6.36  	  	&	0.6	&		3.69 - 5.82       	& 	5.36 - 8.48 \\
Cr2 	&  $\ldots$	& 	0.31 - 0.49				 & 	0.08 - 0.12			& 		0.18 - 0.28		 & 	0.02 - 0.03  	  	&	$\ldots$&		$\ldots$       	&	$\ldots$	 \\
Cr5$^a$ 	&	W10	& 	4.77 - 7.54				 & 	1.25 - 1.97			& 		2.30 - 3.63		 & 	1.74 - 2.76     	  	&	2.8	&		0.82 - 1.29      	&	0.49 - 0.78 \\
Cr5$^b$  &	W10	& 	1.67 - 2.63			 	 & 	0.43 - 0.69			& 		0.88 - 1.40		 & 	0.46 - 0.73  	 		& 	$\ldots$&		$\ldots$        	&	$\ldots$  \\
Cb4 	 &	W45		& 	3.37 - 5.34				 & 	0.88 - 1.40			& 		1.44 - 2.27		 & 	1.08 - 1.71     	 	&	1.9	&		0.76 - 0.89   	&	0.45 - 0.72 \\ 
Cr4 	&	W45		& 	1.90 - 3.01				 & 	0.50 - 0.79 			& 		0.90 - 1.42		 & 	0.70 - 1.11  	  	&	2.1	&		0.43 - 0.67       &	0.26 - 0.42 \\
N1	& 	SMM8	& 	5.09 - 8.05				 &	1.34 -  2.1    			&		1.11 - 1.7		 	& 	0.96 - 1.51     	  	&	1.4	&		0.79 - 1.21	&	0.55 - 0.86	\\
\hline

\hline 
\end{tabular}

~$^a$ extended lobe\\
~$^b$ primary lobe\\
\end{table*}


\section{Outflow Properties}
 \label{properties}
 
In the following sections we calculate the fundamental outflow-related properties (i.e. column densities, masses, momentum, energies. momentum fluxes and mechanical luminosities), employing the method described in \citet{Choi} for the CO \textit{J} = 3$\rightarrow$2 transition. Associated uncertainties are discussed in accordance with the present dataset and in respect with other methods \citep[e.g. ][]{Cabrit1, Bontemps, Fuller, Hatchell2}.
   

\subsection{CO Column Density and Outflow Mass}
\label{mass}

The column density in cm$^{-2}$ per velocity channel \textit{i} for the CO \textit{J} = 3$\rightarrow$2 line was calculated from the relation

\begin{equation}
N^{CO}_{i} =  \frac{3h}{8\pi^3 \mu^2} \frac{T^{\star}_{MB(3-2)}\Delta V}{D(n,T_K)} \frac{\tau_{32}}{1-exp(-\tau_{32})}
\label{eqn:1}
\end{equation} 
 
 where \textit{$\mu$} is the dipole moment of the CO molecule, \textit{T$^{\star}_{MB(3-2)}$} is the corrected mean beam temperature in \textit{K} within the velocity interval \textit{$\Delta V$} (in km s$^{-1}$), $\tau_{32}$ is the optical depth and \textit{D(n,T$_K$)} is given by the relation:
 
 \begin{equation}
 D(n,T) = f_2[J_\nu(T_{ex}) - J_\nu(T_{bk})][1-exp(-h\nu/kT_{ex})]
 \end{equation}
 
 Here, \textit{f$_2$} is the fraction of CO molecules in the \textit{J}=2 state which is dependent on temperature or the excitation conditions in the case of non-LTE, \textit{J(T)} is the Planck function in K,  \textit{T}$_{ex}$ is the excitation temperature and \textit{T}$_{bk}$=2.7K. 
 
The particular characteristic of this method in deriving the column density as opposed to other methods introduced in the literature \citep[e.g.][]{Cabrit1, Bontemps} lies in the treatment of the quantity  \textit{D(n,T)}. \citet{Cabrit1} assumes LTE conditions so that the excitation temperature equals to the kinetic one; at this limit, \textit{D(n,T)} is independent from the density, so that \textit{D(n,T)}=\textit{D(T$_K$)}. Adopted kinetic temperatures in the literature vary from "cold" 10 K \citep[10 - 20  K][]{Cabrit2, Bontemps} to "warm" \citep[$\sim$50 K ][]{Hatchell2} CO gas, depending on the observed transition. \citet{Choi} approach this quantity in more general terms, assuming non-LTE conditions. They employ LVG simulations to test its variance and find that an average value for \textit{D(n,T$_K$)}=1.5 is correct to less than a factor of 2  for the  CO \textit{J}= 3$\rightarrow$2 transition and for temperature ranging between 10 and 200K and density from 10$^4$ to 10$^6$ cm$^{-3}$.

In the absence of isotopic CO observations for the HARP-B map that could provide means for independently estimating the optical depth, we have assumed two values: $\tau$ = 1, and the optical thin limit ($\tau$ = 0). Despite the \textit{J} = 3$\rightarrow$2 CO line in the center is heavily self reversed (see Fig. \ref{fig:spectra}), for the high velocity wings optically thin emission is a reasonable assumption as \citet{White} find for the wings of the \textit{J} = 2$\rightarrow$1 transition a CO/$^{13}$CO $>$ 20. This should be particularly true for \textit{J} = 3$\rightarrow$2 line wings presented here that are more extended to higher velocities and integrated even further from the body of the line. 

We have applied this method summing for all data points enclosed within each of the individual lobes discussed in \S\S \ref{areaA} - \ref{areaC},  Outflows were considered within velocity ranges  -30 km s$^{-1}$ $\leqslant$ V$_{LSR}$ $\leqslant$ 0 km s$^{-1}$ for the blue and 18 km s$^{-1}$ $\leqslant$ V$_{LSR}$ $\leqslant$ 48 km s$^{-1}$ for the red shifted lobes and a velocity step of 1 km s$^{-1}$ has been used. The summed up CO columns for each lobe are reported in column 2 of Table \ref{tab:2}, for the two adopted values of $\tau$.     
    
The mass per channel \textit{Mi} can be derived from the column density per channel (eq. \ref{eqn:1}) given the distance, \textit{D} and the beam solid angle, $\Omega$:

\begin{equation}
M_i = 2\mu m_H D^2 \Omega N^{CO}_{i} \frac{[H]}{[CO]} 
\label{eqn:3}
\end{equation}

where $\mu$ = 1.3 is the mean molecular mass and m$_H$ is the hydrogen atom mass. 
In \citet{Choi} the abundance ratio is separated into two terms with respect to carbon, and two different cases are taken for the [C]/[CO] ratio: one considering CO in swept-up gas and another one treating CO as an intrinsic jet ingredient. Following \citet{Choi}, we consider a  CO abundance ratio \textit{X$_{CO}$} = 5 $\times$ 10$^{-5}$, under the assumption that the observed outflows are indeed swept-up ambient material. The estimated mass for individual lobes and the total mass for the blue and red high velocity gas  considering the two adopted values of $\tau$ are reported in Tables \ref{tab:2} and \ref{tab:3}.

\subsection{Momentum and  Energy}
\label{energy}

Once the mass per channel is calculated for individual outflows, momentum and kinetic energy of each outflow can be derived from the following relations:

\begin{equation}
P = \sum_i M_i \vert V_i - V_0 \vert
\end{equation}

\begin{equation}
E_K = \sum_i \frac{1}{2}M_i  (V_i - V_0)^2
\end{equation}

where \textit{V$_0$} is the nominal cloud velocity and sums act over velocity channels (\textit{i}) reported above for the blue and red-shifted lobes. The derived momentum and kinetic energy for each individual outflows are reported in columns 3 and 4 of Table \ref{tab:2}.  The total mass, momentum and kinetic energy for the high velocity blue and red shifted gas in the Serpens Core are listed in Table \ref{tab:3}; in all cases, lower and upper limits reflect the assumption of the two different values for $\tau$.

The total mass estimates lie between 1.9 and 3.0 $\times$ 10$^{-2}$ M$_\odot$, for the momentum between 3.25 and 5.1 $\times$ 10$^{-1}$ M$_\odot$ km s$^{-1}$ and energy from 4.4-6.9 $\times$ 10$^{43}$ erg. Total values of these properties for the blue-shifted gas are $50\%$ higher than for the red-shifted emission. In comparison to the global outflow mass, momentum and energy for the CO \textit{J} = 2$\rightarrow$1 high velocity gas reported in \citet{White} and \citet{Davis}, our estimations are 1-2 orders of magnitude lower.  For comparison, integrating over the velocity intervals considered in \citet{Davis} 
([-4, 4] km s$^{-1}$ and [14, 22] km s$^{-1}$ for the blue and red-shifted gas) yields 
values which are an order of magnitude lower than the ones derived by the CO \textit{J} = 2$\rightarrow$1. 
Part of this difference could be due to the fact that the area of 720$\arcsec \times$ 560$\arcsec$ 
considered in \citet{Davis} is much larger than the area of the present maps (460$\arcsec \times$ 230$\arcsec$). 
Indeed, considering the velocity limits used in \citet{Graves} ([-15, 6] km s$^{-1}$ and [10, 30] 
km s$^{-1}$ for the blue and red-shifted gas), our estimates are less than a factor of two lower. 
Also the outflow masses and momenta for individual outflows reported in Table 3 of \citet{Graves}, 
are within a factor of two with respect to the ones reported in Table \ref{tab:2}; 
such a difference can be easily due to the different adopted methods for measuring these
quantities.

\subsection{Momentum Flux and Mechanical Luminosity}
\label{momentumflux}

The driving force of the underlying stellar wind or jet should correspond to the average force applied to the outflow gas.  Its magnitude  can be estimated if the momentum of the outflowing gas is divided by the characteristic flow timescale \textit{t$_{ch}$}: 

\begin{equation}
F^{obs}_{CO}=\frac{P}{t_{ch}}
\label{eqn:6}
\end{equation}

The characteristic timescale is defined as the maximum projected radius that the flow has covered \textit{R}, over the mass weighted velocity of the flow \textit{v$_{ch}$ =  P/M}, This timescale represents a lower limit as it is not corrected for any possible inclination of the outflow. In addition, most swept up mass in jet-driven shocks is located in the bow wings and moves much more slowly than material in the bow-head; therefore the mass weighted velocity may contribute in underestimating outflow ages \citep{Downes}. Thus \textit{F$_{CO}$} becomes:

 \begin{equation}
F^{obs}_{CO}=\frac{P^2}{MR}
\label{eqn:7}
\end{equation}   

\begin{table}
\caption{Global properties of the high velocity gas} 
\label{tab:3} 
\centering 
\begin{tabular}{l  c c c} 
\hline\hline 
	&	M$_{tot}$		& 		P$_{tot}$ & E$_{K, tot}$  \\	
        &  (10$^{-2}$ $M_{\sun}$)  & 10$^{-1}$ $M_{\sun}$ km s$^{-1}$) & (10$^{43}$ erg) \\
\hline
Blue  &  1.12 - 1.76		&	1.97 - 3.04	&	2.92 - 4.55		\\ 
Red  &  0.75 - 1.27		&	1.28	- 2.09	& 	1.44	- 2.34  		\\
Sum &  1.87 - 3.03		& 	3.25	- 5.13	& 	4.36	- 6.89 		\\
\hline

\end{tabular}
 \end{table}

Inclination of the outflow with respect to the plane of the sky will lead in underestimating the true outflow length and the characteristic velocity. Adjusting for the inclination effects on the observed momentum flux yields:

 \begin{equation}
F_{CO}=\frac{P^2}{MR} \times \frac{sin (i)}{cos^2 (i)}
\label{eqn:8}
\end{equation}  

where \textit{i} is the angle between the flow axis and the line of sight. As the latter quantity cannot be inferred from the existing data on an individual outflow basis, we have assumed an arbitrary inclination angle \textit{i = 45$^o$} for all outflows corresponding to an inclination correction factor of 1.41. 
In a similar manner, the total energy from the jet deposited on the ambient gas can be derived dividing the outflow energy by the characteristic timescale. 
 
\begin{equation}
L^{obs}_{mech}=\frac{E_K}{t_{ch}}
\label{eqn:8.1}
\end{equation}

The inclination taken into account corresponds to a factor of $sin(i)/cos^{3}(i)$, which for an angle of 45$^o$  results in a correction factor of 2. The derived momentum fluxes and mechanical luminosities for each individual outflow are reported in Table \ref{tab:2}.
 
Results from numerical simulations of outflows \citep{Downes} argue that the characteristic timescale \textit{t$_{ch}$} as defined above will overestimate the true outflow length, as mass weighted velocities underestimate the true propagation speed of the bow-shock head, which defines its actual length in a jet driven outflow. Other factors such as the role of the transverse momentum and the dissociation of CO molecules at bowshocks may affect to a greater or lesser extent the estimations of momentum flux (for a complete discussion we refer the interested reader to \citet{Downes}). However, the derivation of the momentum flux based on the maximum velocity as suggested by \citet{Downes} may lead to more uncertain results, since \textit{v$_{max}$} is more affected by the inclination of the outflow. Further more, this method may give misleading results in the presence of high velocity "bullets", as it is the case in a number of outflows in the present sample. In the current method for deriving the momentum flux, overestimating the true outflow age is compensated to an extent by the high \textit{v$_{min}$} cutoffs adopted for the outflows; for example, assuming v$_{min}$ values closer to v$_{LSR}$ by 4 km s$^{-1}$, the mechanical energy increases by $\sim$ 15 $\%$. Characteristic timescales and momentum fluxes for each outflow are reported in Table \ref{tab:2}. There are two cases where the characteristic timescale could not be determined; for lobe Bb2, the outflow seems to be almost aligned with the line of sight from the driving source W48, while no possible driving source could be assigned to lobe Cr2 (see \S \ref{areaC}).

\begin{figure*}
\centering                             
\resizebox{14cm}{!}{\includegraphics[angle=0]{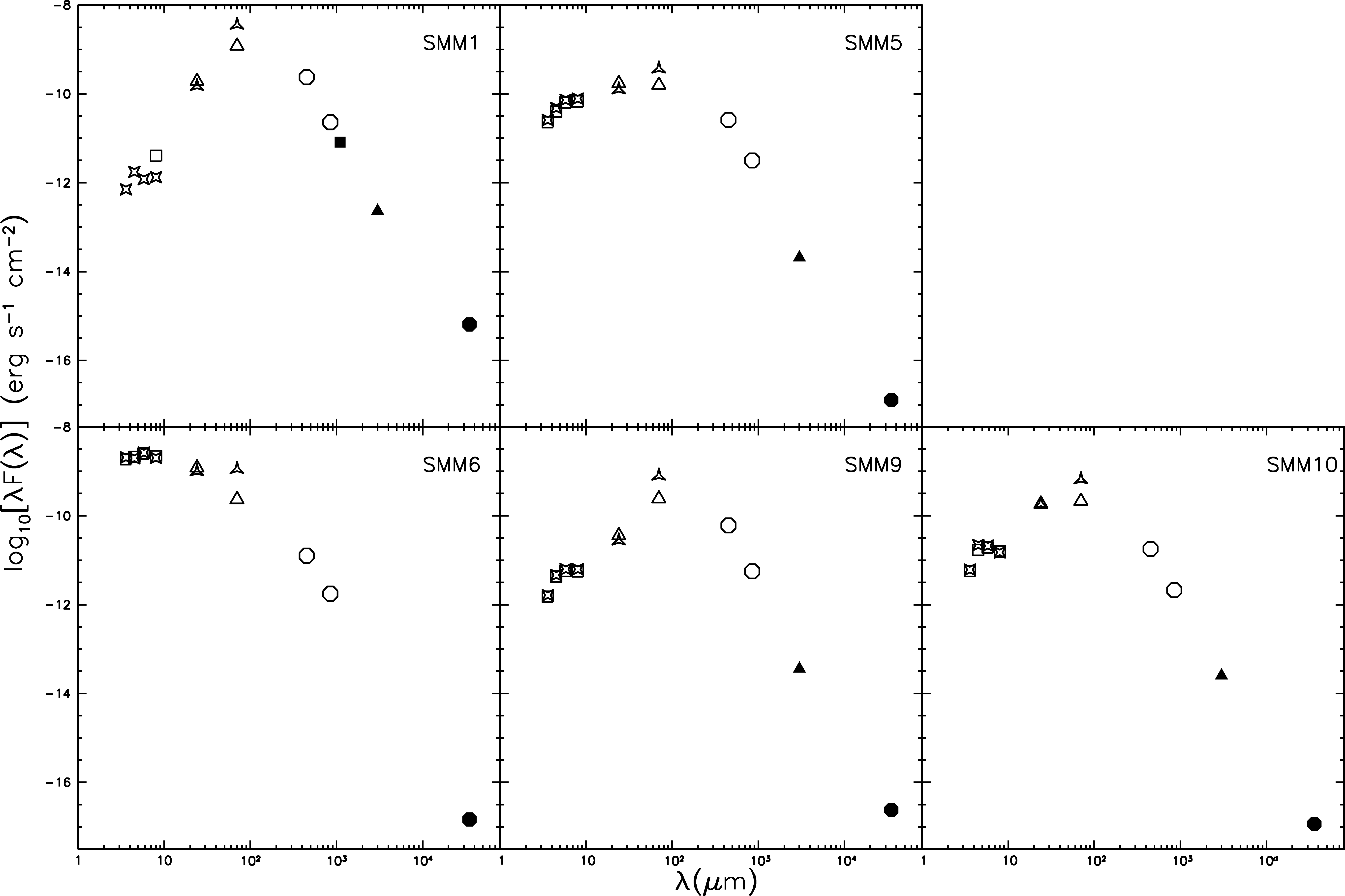}}
\caption{Spectral energy distribution diagrams for the sources exhibiting 70 $\micron$ emission. Empty squares and star like squares are IRAC band 1 to 4 flux densities from \citet{Winston} and \citep{Harvey} respectively; empty triangles and star like triangles are MIPS band 1 and 2 from  \citet{Winston} and \citep{Harvey}. Empty circles are sub-mm flux densities at 450 and 850 $\micron$ from \citet{Davis}. Filled squares and triangles are 1.1 and 3 millimeter flux densities from \citet{Enoch1} and \citet{Williams}. Filled circles are 3.5 centimeter flux densities from \citet{Eiroa}. Note that the 70 $\micron$ flux of \citet{Harvey} is constantly higher by a factor $\sim$2-3 compared to the one reported by \citet{Winston}}
\label{fig:sed}
\end{figure*}

There are no studies where CO momentum fluxes are estimated for individual sources in Serpens. However comparison with other studies shows that  derived values presented in Table \ref{tab:2} are of the same order of magnitude as the ones in the case of \citet{Bontemps} examining a sample of 45 low-luminosity embedded young stellar objects, and 2 orders of magnitude higher in respect to the lower limits inferred by \citet{Hatchell2} for a sample of embedded sources in Perseus cloud. There is a factor of $\sim$ 10 implied by the opacity and inclination corrections implemented in the momentum fluxes of \citet{Bontemps} that are not taken into account by \citet{Hatchell2}. Even corrected for this factor,  there is a discrepancy of an order of magnitude even for the outflows attributed to the same sources in the two studies, which reflects the high uncertainties in the momentum flux calculations.

Based on the observed CS outflow associated with S68N/SMM9 core, \citet{Wolf-Chase} estimated a momentum flux F$_{CO} \sim$ 2 $\times$ 10 $^{-4}$ M$_{\sun}$ km s$^{-1}$ yr$^{-1}$; this is an order of magnitude higher than the sum of the lobes Ar1 and Ab1, which are attributed to SMM9 in our study. Even though the SMM9 outflow being at the edge of out map is not fully sampled, the value derived by \citet{Wolf-Chase} is still high as it corresponds to the total momentum flux for the whole Serpens core in the optically thin limit ($\sim$ 2.8 - 5 $\times$ 10$^{-4}$ M$_{\sun}$ km s$^{-1}$ yr$^{-1}$). A series of factors including the assumed dynamical timescale along with the low velocity cutoffs (which go much into the body of the CS lines in \citet{Wolf-Chase}) may most easily explain this discrepancy. It should also be noted that the relatively high CS abundance employed in \citet{Wolf-Chase} is expected to lower the momentum flux (see eq. \ref{eqn:3}).  However, the CS abundance is highly uncertain and thus the comparison between CO and CS derived momentum fluxes is not straightforward. The total momentum flux estimated by \citet{Davis} for all the outflows in Serpens, ($\sim$3 $\times$ 10$^{-4}$ M$_{\sun}$ km s$^{-1}$ yr$^{-1}$) is in good agreement with our measurements. However, in this case, the $\sim$10 times higher estimated outflow momentum (see \S \ref{energy}) in \citet{Davis} is compensated by  an equally overestimated average characteristic timescale (3 $\times$ 10$^4$ yr), due to the low values of the assumed maximum gas velocity and the extended typical length  estimated for the outflows.   

\subsection{Bolometric Luminosity}
\label{luminosity}
The spectral energy distribution (SED) of very young, embedded Class0/I sources spans in wavelength a range from the near or mid-IR to mm and cm wavelengths; typically, the observed flux density is low in the near and mid-IR, while it peaks in the far-IR as the reprocessed protostellar flux is emitted by the massive surrounding envelope, falling back to lower values at longer wavelengths \citep{Enoch2}. 

Estimations of the bolometric luminosity for the very young protostars in the Serpens core differ substantially in diverse studies \citep{Casali, Hurt1, Larsson}. In particular, the large beams of the mid and far-IR facilities deployed (e.g. \textit{IRAS}; \citet{Hurt1} and \textit{ISO}; \citet{Larsson}), do not posses enough spatial resolving power to resolve the emissions from the highly clustered YSOs in the Serpens core. This leads to overestimated SED peak values for very young protostars and consequently to higher  bolometric luminosities. Indeed, recent calculations based on \textit{Spitzer} far-IR data which have an improved spatial resolution \citep{Enoch2}, find \textit{L$_{bol}$} values up to an order of magnitude lower than previous studies.

\begin{table*}
\caption{Momentum flux and bolometric luminosity} 
\label{tab:4}
\centering  
\begin{tabular}{l c c c c c} 
\hline\hline 
Source	&	L$_{bol}^{a}$ & 	 M$_{CO}$ & F$_{CO}^b$ & L$_{mech}^b$ & L$_{smm}$/L$_{bol}$	\\	
	& 	(L$_{\sun}$)	&	(10$^{-3}$ M$_{\sun}$) &	(10$^{-5}$ $M_{\sun}$ km s$^{-1}$ yr$^{-1}$) & (10$^{-2}$ L$_{\sun}$) & (10$^{-2}$)	 \\
\hline

SMM1 & 7.7    $\pm$ 2.4  	&   6.45 $\pm$ 1.45  &  6.48 $\pm$ 3.24   &     12.96 $\pm$ 2.92  & 4.4 $\pm$ 0.9 	\\ 
SMM3 & 1.05  $\pm$ 0.5   	&   1.33 $\pm$ 0.3   &  3.04 $\pm$ 1.23   &      4.42 $\pm$ 1.0   & 5.8 $\pm$ 1.1     \\
SMM4 & 0.25  $\pm$ 0.05 	&   3.2 $\pm$ 0.72   &  9.03 $\pm$ 2.01   &      12.24 $\pm$ 2,74   & 4.8 $\pm$ 0.8     \\
SMM5 & 1.62  $\pm$ 0.22 	&   0.89 $\pm$ 0.19  &  0.47 $\pm$ 0.11   &     0.46 $\pm$ 0.1       &  0.3 $\pm$ 0.1 \\
SMM6 & 13.54 $\pm$ 0.8  	&   1.52 $\pm$ 0.34  &  6.71 $\pm$ 1.5    &      13.84 $\pm$ 3.12     & 0.2 $\pm$ 0.1   \\
SMM8 & 0.068   			&   1.72 $\pm$ 0.38  &  1.41 $\pm$ 0.29   &     1.4 $\pm$ 0.3         &  $\cdots$\\
SMM9 & 1.73  $\pm$ 0.3 		&   1.26 $\pm$ 0.29  &  4.43 $\pm$ 1.28   &     5.34 $\pm$ 1.22      & 5.8 $\pm$ 0.7  \\
W45  & 0.027   				&   1.78 $\pm$ 0.40  &  1.95 $\pm$ 0.26   &     1.84 $\pm$ 0.42      & $\cdots$  \\
W10  & 2.10  $\pm$ 0.1  		&   2.17 $\pm$ 0.49  &  1.49 $\pm$ 0.33   &     1.26 $\pm$ 0.28      & $\cdots$ \\
S68Nb& 0.47     			&   0.95 $\pm$ 0.35  &  0.55 $\pm$ 0.12   &     0.46 $\pm$ 0.1        & 0.2 $\pm$ 0.1\\

\hline
\end{tabular}

~ $^a$ lower and upper limits based on Spitzer data from \citet{Winston} and  \citet{Harvey}\\
~$^b$ values corrected for inclination angle \textit{i} = 45$^o$
\end{table*}

Here, we estimate bolometric luminosities for the sources that are likely to be responsible for the observed outflow activity, compiling data extending from the mid and far-IR \citep{Winston, Harvey}, 450 and 850 $\micron$ in the sub-millimeter \citep{Davis}, 1.1 and 3 millimeter \citet{Enoch1} and\citet{Williams} to 3.5 cm \citet{Eiroa} observations of the Serpens core.  The Spitzer mid and far-IR photometry reported in the two independent studies of \citet{Winston} and \citet{Harvey} are in good agreement except for the 70 $\micron$ MIPS band, where values reported by \citet{Harvey} are always higher by a factor between 2 and 3. As the bulk of the derived bolometric luminosity rises from the area where flux density peaks (i.e. the far-IR for Class 0 sources), the 70 $\micron$ flux has a dominant role. We therefore estimate upper and lower L$_{bol}$ limits, taking into account the two independent Spitzer datasets. Spectral energy distribution diagrams for the \citet{Winston} catalogue sources exhibiting 70 $\micron$ emission along with flux densities estimated in \citet{Harvey} are presented in Figure \ref{fig:sed}. 
  
Flux density in the SEDs is sampled over a wide range of wavelengths with a small collection of data points; before integrating, we use a simple linear interpolation method similar to the "mid-point" method employed by \citet{Enoch2}. However, while in the "mid-point" method a single interpolant is used, in our calculations instead, interpolants are introduced over a constant wavelength step of 5 $\micron$. Upper and lower limit bolometric luminosities  obtained with this method for the adopted distance of  310 pc are reported  Table \ref{tab:4}. 

The derived bolometric luminosities are in good agreement with the ones reported in \citet{Enoch2}, despite the different assumptions and methods. L$_{bol}$ upper and lower limits in \citet{Enoch2} are estimated by means of two different integration methods (see Appendix A in \citet{Enoch2}) considering only the \textit{c2d} survey dataset of \citet{Harvey} and assuming a 260 pc distance to Serpens. It should be noted however that peak fluxes are interpolated over in the "mid-point" method, and therefore the dominant influence of the 70 $\micron$ band is lost. Thus in principle, bolometric luminosities estimated here and in the case of \citet{Enoch2} should differ by a factor of (310)$^2$/(260)$^2$ $\sim$ 1.4 due to the different adopted distances. Thorough tests on the same dataset of the two integration methods have shown that the predicted bolometric luminosities depend strongly on the number of interpolants between the observed points, and the estimated values are lower by a factor of $\sim$ 1.5 compared to the single interpolant adopted by \citet{Enoch2}, canceling the difference in the adopted distances. 

In Table \ref{tab:4} we also report the ratio of the submillimeter ($>$ 350 $\micron$) to the bolometric luminosity, for the sources having enough observed sub-millimeter and millimeter data points . This ratio being greater than 0.5 $\%$ is indicative of the envelope mass exceeding the central stellar mass, therefore distinguishing between Class 0 and more evolved protostars \citep{Andre}. Associated errors are estimated considering the different values reported in the two mid/far IR studies with \textit{Spitzer} \citep{Winston, Harvey}. Four protostellar sources, namely SMM1, SMM3 SMM4 and SMM9 fulfill this criterion and therefore are candidate Class 0 protostars in Serpens. We stress out however that this ratio is  dependent on the physical and observed properties of a particular source such as the inclination of outflow cavity with respect to the line-of-sight, therefore it only may provide indications and not constraints on the evolutionary stage of a protostellar source.

\section{Discussion}\label{discussion}
\subsection{Outflow Momentum Flux and Mechanical Luminosity versus Bolometric Luminosity relations}
\label{correlation}

\begin{figure}                             
\centering
\resizebox{8cm}{!}{\includegraphics{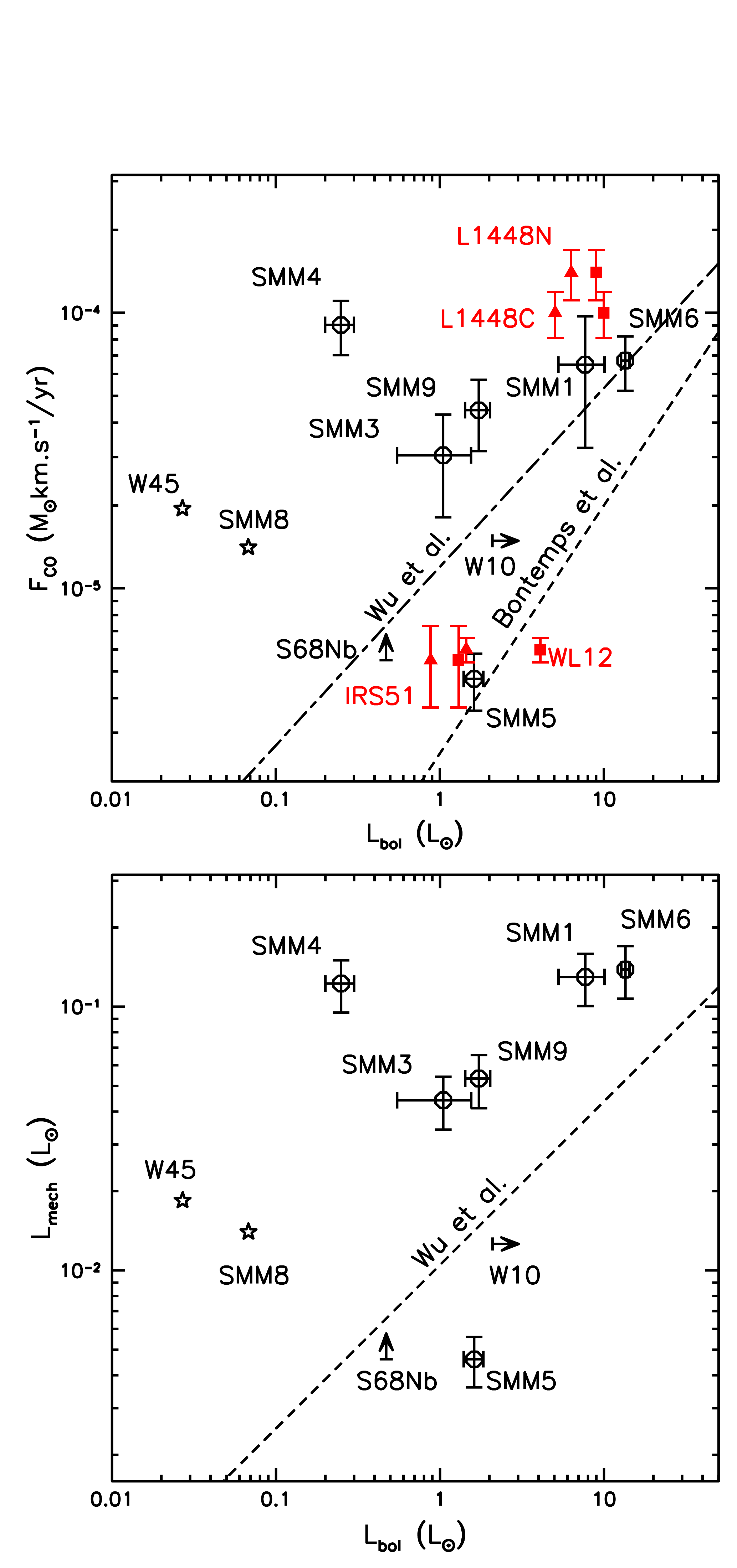}}
\caption{(\textit{Upper panel}):Momentum flux vs bolometric luminosity. Open circles indicate sources from the current sample of Serpens; vertical errorbars correspond to values of the momentum flux for $\tau$ = 0 and 1, and horizontal errorbars to the differences in the derived bolometric luminosities for the different Spitzer fluxes in \citet{Winston} and Harvey. Arrows and asterisks mark Serpens sources that either F$_{CO}$, L$_{bol}$ or both are uncertain. The dashed   line shows the location of the Class I correlation of \citet{Bontemps} and the dot-dashed one the correlation of both low and high mass sources of \citep{Wu}. Filled (red) squares correspond to test sources from the sample \citet{Bontemps}, used to control the influence of the new Spitzer derived (red triangles) L$_{bol}$ to the correlation. (\textit{Lower panel}):  Outflow mechanical luminosity veersus the bolometric luminosity for the sample of sources in Serpens; the L$_{mech}$ - L$_{bol}$ correlation for the sample of both low and high mass sources of \citet{Wu} is shown as a dashed line, running through the data points and separating Class 0 and Class I sources. }  
\label{fig:fl}
\end{figure}

We examine the well known correlation between momentum flux and bolometric luminosity that has been proposed by \citep{Cabrit2} and \citet{Bontemps} for the homogenous sample of young protostellar objects in the Serpens Core. The outflow momentum flux \textit{F$_{CO}$} is plotted against the bolometric luminosity \textit{L$_{bol}$} in the upper panel of  Figure \ref{fig:fl}. Empty circles correspond to Serpens sources with associated outflows that both quantities could be determined, with error bars displaying the corresponding uncertainties discussed n the previous sections. Arrows and asterisks correspond to sources/outflows that either \textit{F$_{CO}$},  \textit{L$_{bol}$} or both are uncertain. The dashed and dot-dashed lines display the location for the Class I  correlation of \citet{Bontemps} and the relation for a sample of both low and high mass protostellar sources of \citep{Wu}.

All protostellar sources found to have a high submillimeter to bolometric luminosity ratios stand well above of both \citet{Bontemps} and \citep{Wu} correlations.
In addition, Class 0 and I sources are distributed above and below the \citet{Wu} correlation, which derived for a large sample of both low and high mass protostars. The placement of Class 0 sources confirms the soundness of our estimations, as the  \citep{Wu} relation is strongly affected by high mass protostars, which are expected to have short evolutionary times and therefore are expected to be older than the low mass Class 0 protostars. 

While the sample of Class I sources is not large enough to drive statistically significant conclusions, it appears that Class I sources are shifted to lower luminosities with respect to what is predicted from the correlation of \citet{Bontemps}. This shift should not come as a surprise, since bolometric luminosities reported in this work and in \citet{Enoch2} are up to a factor of $\sim$ 10 lower than previous estimations \citep{Casali, Hurt1, Larsson}, this difference being attributed to  the better spatial resolution of \textit{Spitzer}. 

In order to explore the effect of the updated  bolometric luminosities on the Class I correlation, we have selected four additional sources from the \citet{Bontemps} sample, since the latter contains no sources belonging to Serpens that could be used for direct comparison. From these, L1448C and N are bona-fide Class 0 protostars in Perseus and their SEDs were determined using Spitzer photometry \citep{Jorgensen, Rebull} and (sub)-mm observations \citep{Hatchell1, Enoch3}. The other two (IRS51 and WL12), are characterized as Class I sources \citep{Bontemps} located in Ophiuchus, and for their SEDs we have compiled observations from Spitzer \citep[c2d database,][]{Evans, Padgett}, and (sub)-mm \citep{Johnstone, Stanke}. In Figure \ref{fig:fl}, data points for these four sources as originally reported in \citet{Bontemps} (filled squares) are brighter by a factor between 2 and 3 than the bolometric luminosities based on Spitzer data (filled triangles). Despite this shift, Class I sources IRS51 and WL12 are roughly in the same part of the diagram with Class I candidates from the Serpens sample; the corresponding shift for the Class 0 sources places them even further from the Class I correlation, so that a decline in momentum flux from the Class 0 to the Class I phase is preserved. In another, uniform sample survey of Perseus, \citet{Hatchell2} found a less clear separation for the two classes; however, in for the Serpens sample, all confirmed Class 0 sources seem to be standing well off the Class I group. With the exception of SMM4, all SMM1, 3, and 9 together with Perseus Class 0 sources seem to be well aligned indicating higher CO thrust than the Class I case (an order of magnitude for L$_{bol}$ = 1 L$_\odot$). Hints of a possible alignment between these Class 0  sources exists, however a larger sample of protostellar sources would be necessary to confirm it. To this direction, the HARP-B CO \textit{J} = 3$\rightarrow$2 maps of Perseus \citep{Hatchell3} along with the existing \textit{Spitzer} and (sub)-mm surveys of the region combined with the present data would be an ideal dataset for probing this possibility.

Extending the investigation of the relations between outflows and the driving sources in Serpens,  in the lower panel of Fig. \ref{fig:fl} we plot the mechanical luminosity (L$_{mech}$) of the outflows versus the bolometric luminosity (L$_{bol}$) of the corresponding protostellar sources. The dashed line shows the corresponding correlation of \citet{Wu} comprising  both low and high mass protostars. The distribution of data points is similar to the F$_{CO}$ vs L$_{bol}$ in the upper panel of the same Figure and is consistent with the data distribution of \citet{Wu}. Similarly, Class 0 sources stand well above the \citet{Wu} correlation; this is not unexpected, as the \citet{Wu} correlation is derived for a large sample of sources spanning different masses and ages. Despite the sample is limited, it appears that sources with high bolometric luminosities tend to produce outflows with relatively lower mechanical luminosity in respect with the sources having lower bolometric luminosities;  a similar behavior may be also observed in the sample of \citep{Wu}.

\section{Summary and Conclusions}
\label{conclusions}

We have carried out CO \textit{J} = 3$\rightarrow$2 observations of the Serpens Cloud Core with the HARP-B detector on JCMT, covering an area of 460$\arcsec\times$230$\arcsec$. These higher \textit{J} CO observations along with the higher sensitivity of the HARP-B have allowed to sharply delineate  the outflow activity in Serpens, a fact that has enabled to a greater extent than before the association of individual outflows with protostellar sources. Such associations were aided by comparing with other jet/outflow tracers, performed in other studies in the literature, and have resulted in associating most observed outflows to $\sim$10 protostellar sources. The projected orientation on the plane of the sky for the majority of the outflows, displays a preferable NW-SE direction. This orientation is roughly parallel to the local galactic magnetic field as predicted by magnetically controlled could collapse models. For individual outflows, deduced column densities, masses as well as kinematical and dynamical properties are in good agreement with previous studies. However, estimated bolometric luminosities based on Spitzer mid and far-IR observations are lower up to a factor of 10 than previous estimations based on IRAS/ISO data. This difference is reflected as a possible shift in the F$_{CO}$ vs L$_{bol}$ Class I correlation of \citep{Bontemps} to lower luminosities. Still, the distribution of Class I sources remains well separated from the one of Class 0 protostars. Indications for an independent F$_{CO}$ - L$_{bol}$ correlation for the latter sources exist in the present sample, however its limited extent is not sufficient to drive secure conclusions.

\begin{acknowledgements} 
This work is based on observations made with the James Clerk Maxwell Telescope, operated by the Joint Astronomy Centre on behalf of the Science and Technology Facilities Council of the U.K., and was supported in part by the European Community's Marie Curie Actions - Human Resource and Mobility within the JETSET (Jet Simulations, Experiments and Theory) network under contract MRTN-CT-2004 05592. The Centre for Star and Planet Formation is funded by the Danish National Research Foundation and the University of Copenhagen's programme of excellence.
\end{acknowledgements}

\bibliographystyle{aa} 
\bibliography{serpens} 

\begin{thebibliography}{62}
\expandafter\ifx\csname natexlab\endcsname\relax\def\natexlab#1{#1}\fi

\bibitem[{{Andre} {et~al.}(1993){Andre}, {Ward-Thompson}, \& {Barsony}}]{Andre}
{Andre}, P., {Ward-Thompson}, D., \& {Barsony}, M. 1993, \apj, 406, 122

\bibitem[{{Bachiller} {et~al.}(1991){Bachiller}, {Martin-Pintado}, \&
  {Fuente}}]{Bachiller2}
{Bachiller}, R., {Martin-Pintado}, J., \& {Fuente}, A. 1991, \aap, 243, L21

\bibitem[{{Bontemps} {et~al.}(1996){Bontemps}, {Andre}, {Terebey}, \&
  {Cabrit}}]{Bontemps}
{Bontemps}, S., {Andre}, P., {Terebey}, S., \& {Cabrit}, S. 1996, \aap, 311,
  858

\bibitem[{{Cabrit} \& {Bertout}(1990)}]{Cabrit1}
{Cabrit}, S. \& {Bertout}, C. 1990, \apj, 348, 530

\bibitem[{{Cabrit} \& {Bertout}(1992)}]{Cabrit2}
{Cabrit}, S. \& {Bertout}, C. 1992, \aap, 261, 274

\bibitem[{{Cambr{\'e}sy}(1999)}]{Cambresy}
{Cambr{\'e}sy}, L. 1999, \aap, 345, 965

\bibitem[{{Casali} {et~al.}(1993){Casali}, {Eiroa}, \& {Duncan}}]{Casali}
{Casali}, M.~M., {Eiroa}, C., \& {Duncan}, W.~D. 1993, \aap, 275, 195

\bibitem[{{Choi} {et~al.}(1993){Choi}, {Evans}, \& {Jaffe}}]{Choi}
{Choi}, M., {Evans}, II, N.~J., \& {Jaffe}, D.~T. 1993, \apj, 417, 624

\bibitem[{{Curiel} {et~al.}(1993){Curiel}, {Rodriguez}, {Moran}, \&
  {Canto}}]{Curiel1}
{Curiel}, S., {Rodriguez}, L.~F., {Moran}, J.~M., \& {Canto}, J. 1993, \apj,
  415, 191

\bibitem[{{Davis} {et~al.}(1999){Davis}, {Matthews}, {Ray}, {Dent}, \&
  {Richer}}]{Davis}
{Davis}, C.~J., {Matthews}, H.~E., {Ray}, T.~P., {Dent}, W.~R.~F., \& {Richer},
  J.~S. 1999, \mnras, 309, 141

\bibitem[{{de Lara} {et~al.}(1991){de Lara}, {Chavarria-K.}, \&
  {Lopez-Molina}}]{deLara}
{de Lara}, E., {Chavarria-K.}, C., \& {Lopez-Molina}, G. 1991, \aap, 243, 139

\bibitem[{{Dent} {et~al.}(2000){Dent}, {Duncan}, {Ellis}, {Harris},
  {Lightfoot}, {Wall}, {Gibson}, {Hills}, {Richer}, {Smith}, {Withington},
  {Burgess}, {Casorso}, {Dewdney}, {Hovey}, {Redman}, {Yeung}, {Force}, \&
  {Pain}}]{Dent}
{Dent}, W., {Duncan}, W., {Ellis}, M., {et~al.} 2000, in Astronomical Society
  of the Pacific Conference Series, Vol. 217, Imaging at Radio through
  Submillimeter Wavelengths, ed. J.~G. {Mangum} \& S.~J.~E. {Radford}, 33--+

\bibitem[{{Downes} \& {Cabrit}(2007)}]{Downes}
{Downes}, T.~P. \& {Cabrit}, S. 2007, \aap, 471, 873

\bibitem[{{Eiroa} \& {Casali}(1992)}]{Eiroa3}
{Eiroa}, C. \& {Casali}, M.~M. 1992, \aap, 262, 468

\bibitem[{{Eiroa} {et~al.}(1997){Eiroa}, {Palacios}, {Eisloffel}, {Casali}, \&
  {Curiel}}]{Eiroa2}
{Eiroa}, C., {Palacios}, J., {Eisloffel}, J., {Casali}, M.~M., \& {Curiel}, S.
  1997, in IAU Symposium, Vol. 182, Herbig-Haro Flows and the Birth of Stars,
  ed. B.~{Reipurth} \& C.~{Bertout}, 103P--+

\bibitem[{{Eiroa} {et~al.}(2005){Eiroa}, {Torrelles}, {Curiel}, \&
  {Djupvik}}]{Eiroa}
{Eiroa}, C., {Torrelles}, J.~M., {Curiel}, S., \& {Djupvik}, A.~A. 2005, \aj,
  130, 643

\bibitem[{{Enoch} {et~al.}(2008){Enoch}, {Evans}, {Sargent}, \&
  {Glenn}}]{Enoch2}
{Enoch}, M.~L., {Evans}, II, N.~J., {Sargent}, A.~I., \& {Glenn}, J. 2008,
  ArXiv e-prints

\bibitem[{{Enoch} {et~al.}(2007){Enoch}, {Glenn}, {Evans}, {Sargent}, {Young},
  \& {Huard}}]{Enoch1}
{Enoch}, M.~L., {Glenn}, J., {Evans}, II, N.~J., {et~al.} 2007, \apj, 666, 982

\bibitem[{{Enoch} {et~al.}(2006){Enoch}, {Young}, {Glenn}, {Evans}, {Golwala},
  {Sargent}, {Harvey}, {Aguirre}, {Goldin}, {Haig}, {Huard}, {Lange},
  {Laurent}, {Maloney}, {Mauskopf}, {Rossinot}, \& {Sayers}}]{Enoch3}
{Enoch}, M.~L., {Young}, K.~E., {Glenn}, J., {et~al.} 2006, \apj, 638, 293

\bibitem[{{Evans} {et~al.}(2003){Evans}, {Allen}, {Blake}, {Boogert}, {Bourke},
  {Harvey}, {Kessler}, {Koerner}, {Lee}, {Mundy}, {Myers}, {Padgett},
  {Pontoppidan}, {Sargent}, {Stapelfeldt}, {van Dishoeck}, {Young}, \&
  {Young}}]{Evans}
{Evans}, II, N.~J., {Allen}, L.~E., {Blake}, G.~A., {et~al.} 2003, \pasp, 115,
  965

\bibitem[{{Fuller} \& {Ladd}(2002)}]{Fuller}
{Fuller}, G.~A. \& {Ladd}, E.~F. 2002, \apj, 573, 699

\bibitem[{{Garay} {et~al.}(2002){Garay}, {Mardones}, {Rodr{\'{\i}}guez},
  {Caselli}, \& {Bourke}}]{Garay}
{Garay}, G., {Mardones}, D., {Rodr{\'{\i}}guez}, L.~F., {Caselli}, P., \&
  {Bourke}, T.~L. 2002, \apj, 567, 980

\bibitem[{{Giovannetti} {et~al.}(1998){Giovannetti}, {Caux}, {Nadeau}, \&
  {Monin}}]{Giovannetti}
{Giovannetti}, P., {Caux}, E., {Nadeau}, D., \& {Monin}, J.-L. 1998, \aap, 330,
  990

\bibitem[{{Gomez de Castro} {et~al.}(1988){Gomez de Castro}, {Eiroa}, \&
  {Lenzen}}]{Gomez}
{Gomez de Castro}, A.~I., {Eiroa}, C., \& {Lenzen}, R. 1988, \aap, 201, 299

\bibitem[{{Graves} {et~al.}(2010){Graves}, {Richer}, {Buckle}, {Duarte-Cabral},
  {Fuller}, {Hogerheijde}, {Owen}, {Brunt}, {Butner}, {Cavanagh},
  {Chrysostomou}, {Curtis}, {Davis}, {Etxaluze}, {Di Francesco}, {Friberg},
  {Friesen}, {Greaves}, {Hatchell}, {Johnstone}, {Matthews}, {Matthews},
  {Matzner}, {Nutter}, {Rawlings}, {Roberts}, {Sadavoy}, {Simpson}, {Tothill},
  {Tsamis}, {Viti}, {Ward-Thompson}, {White}, {Wouterloot}, \&
  {Yates}}]{Graves}
{Graves}, S.~F., {Richer}, J.~S., {Buckle}, J.~V., {et~al.} 2010, ArXiv
  e-prints

\bibitem[{{Gregersen} {et~al.}(1997){Gregersen}, {Evans}, {Zhou}, \&
  {Choi}}]{Gregersen}
{Gregersen}, E.~M., {Evans}, II, N.~J., {Zhou}, S., \& {Choi}, M. 1997, \apj,
  484, 256

\bibitem[{{Haisch} {et~al.}(2002){Haisch}, {Barsony}, {Greene}, \&
  {Ressler}}]{Haisch}
{Haisch}, Jr., K.~E., {Barsony}, M., {Greene}, T.~P., \& {Ressler}, M.~E. 2002,
  \aj, 124, 2841

\bibitem[{{Harvey} {et~al.}(2007){Harvey}, {Mer{\'{\i}}n}, {Huard}, {Rebull},
  {Chapman}, {Evans}, \& {Myers}}]{Harvey}
{Harvey}, P., {Mer{\'{\i}}n}, B., {Huard}, T.~L., {et~al.} 2007, \apj, 663,
  1149

\bibitem[{{Hatchell} \& {Dunham}(2009)}]{Hatchell3}
{Hatchell}, J. \& {Dunham}, M.~M. 2009, ArXiv e-prints

\bibitem[{{Hatchell} {et~al.}(2007{\natexlab{a}}){Hatchell}, {Fuller}, \&
  {Richer}}]{Hatchell2}
{Hatchell}, J., {Fuller}, G.~A., \& {Richer}, J.~S. 2007{\natexlab{a}}, \aap,
  472, 187

\bibitem[{{Hatchell} {et~al.}(2007{\natexlab{b}}){Hatchell}, {Fuller},
  {Richer}, {Harries}, \& {Ladd}}]{Hatchell1}
{Hatchell}, J., {Fuller}, G.~A., {Richer}, J.~S., {Harries}, T.~J., \& {Ladd},
  E.~F. 2007{\natexlab{b}}, \aap, 468, 1009

\bibitem[{{Herbst} {et~al.}(1997){Herbst}, {Beckwith}, \& {Robberto}}]{Herbst}
{Herbst}, T.~M., {Beckwith}, S.~V.~W., \& {Robberto}, M. 1997, \apjl, 486, L59+

\bibitem[{{Hodapp}(1999)}]{Hodapp}
{Hodapp}, K.~W. 1999, \aj, 118, 1338

\bibitem[{{Hogerheijde} {et~al.}(1999){Hogerheijde}, {van Dishoeck},
  {Salverda}, \& {Blake}}]{Hogerheijde}
{Hogerheijde}, M.~R., {van Dishoeck}, E.~F., {Salverda}, J.~M., \& {Blake},
  G.~A. 1999, \apj, 513, 350

\bibitem[{{Huard} {et~al.}(1997){Huard}, {Weintraub}, \& {Kastner}}]{Huard}
{Huard}, T.~L., {Weintraub}, D.~A., \& {Kastner}, J.~H. 1997, \mnras, 290, 598

\bibitem[{{Hurt} \& {Barsony}(1996)}]{Hurt1}
{Hurt}, R.~L. \& {Barsony}, M. 1996, \apjl, 460, L45+

\bibitem[{{Johnstone} {et~al.}(2000){Johnstone}, {Wilson}, {Moriarty-Schieven},
  {Joncas}, {Smith}, {Gregersen}, \& {Fich}}]{Johnstone}
{Johnstone}, D., {Wilson}, C.~D., {Moriarty-Schieven}, G., {et~al.} 2000, \apj,
  545, 327

\bibitem[{{J{\o}rgensen} {et~al.}(2006){J{\o}rgensen}, {Harvey}, {Evans},
  {Huard}, {Allen}, {Porras}, {Blake}, {Bourke}, {Chapman}, {Cieza}, {Koerner},
  {Lai}, {Mundy}, {Myers}, {Padgett}, {Rebull}, {Sargent}, {Spiesman},
  {Stapelfeldt}, {van Dishoeck}, {Wahhaj}, \& {Young}}]{Jorgensen}
{J{\o}rgensen}, J.~K., {Harvey}, P.~M., {Evans}, II, N.~J., {et~al.} 2006,
  \apj, 645, 1246

\bibitem[{{Kaas}(1999)}]{Kaas1}
{Kaas}, A.~A. 1999, \aj, 118, 558

\bibitem[{{Kaas} {et~al.}(2004){Kaas}, {Olofsson}, {Bontemps}, {Andr{\'e}},
  {Nordh}, {Huldtgren}, {Prusti}, {Persi}, {Delgado}, {Motte}, {Abergel},
  {Boulanger}, {Burgdorf}, {Casali}, {Cesarsky}, {Davies}, {Falgarone},
  {Montmerle}, {Perault}, {Puget}, \& {Sibille}}]{Kaas2}
{Kaas}, A.~A., {Olofsson}, G., {Bontemps}, S., {et~al.} 2004, \aap, 421, 623

\bibitem[{{Larsson} {et~al.}(2000){Larsson}, {Liseau}, {Men'shchikov},
  {Olofsson}, {Caux}, {Ceccarelli}, {Lorenzetti}, {Molinari}, {Nisini},
  {Nordh}, {Saraceno}, {Sibille}, {Spinoglio}, \& {White}}]{Larsson}
{Larsson}, B., {Liseau}, R., {Men'shchikov}, A.~B., {et~al.} 2000, \aap, 363,
  253

\bibitem[{{Lightfoot} {et~al.}(2000){Lightfoot}, {Dent}, {Willis}, \&
  {Hovey}}]{Lightfoot}
{Lightfoot}, J.~F., {Dent}, W.~R.~F., {Willis}, A.~G., \& {Hovey}, G.~J. 2000,
  in Astronomical Society of the Pacific Conference Series, Vol. 216,
  Astronomical Data Analysis Software and Systems IX, ed. N.~{Manset},
  C.~{Veillet}, \& D.~{Crabtree}, 502--+

\bibitem[{{Mangum} {et~al.}(1996){Mangum}, {Latter}, \& {McMullin}}]{Mangum}
{Mangum}, J.~G., {Latter}, W.~B., \& {McMullin}, J.~P. 1996, in IAU Symposium,
  Vol. 170, CO: Twenty-Five Years of Millimeter-Wave Spectroscopy, ed. W.~B.
  {Latter}, J.~E. {Radford Simon}, P.~R. {Jewell}, J.~G. {Mangum}, \&
  J.~{Bally}, 76P--+

\bibitem[{{Narayanan} {et~al.}(2002){Narayanan}, {Moriarty-Schieven}, {Walker},
  \& {Butner}}]{Narayanan}
{Narayanan}, G., {Moriarty-Schieven}, G., {Walker}, C.~K., \& {Butner}, H.~M.
  2002, \apj, 565, 319

\bibitem[{{Padgett} {et~al.}(2008){Padgett}, {Rebull}, {Stapelfeldt},
  {Chapman}, {Lai}, {Mundy}, {Evans}, {Brooke}, {Cieza}, {Spiesman},
  {Noriega-Crespo}, {McCabe}, {Allen}, {Blake}, {Harvey}, {Huard},
  {J{\o}rgensen}, {Koerner}, {Myers}, {Sargent}, {Teuben}, {van Dishoeck},
  {Wahhaj}, \& {Young}}]{Padgett}
{Padgett}, D.~L., {Rebull}, L.~M., {Stapelfeldt}, K.~R., {et~al.} 2008, \apj,
  672, 1013

\bibitem[{{Rebull} {et~al.}(2007){Rebull}, {Stapelfeldt}, {Evans},
  {J{\o}rgensen}, {Harvey}, {Brooke}, {Bourke}, {Padgett}, {Chapman}, {Lai},
  {Spiesman}, {Noriega-Crespo}, {Mer{\'{\i}}n}, {Huard}, {Allen}, {Blake},
  {Jarrett}, {Koerner}, {Mundy}, {Myers}, {Sargent}, {van Dishoeck}, {Wahhaj},
  \& {Young}}]{Rebull}
{Rebull}, L.~M., {Stapelfeldt}, K.~R., {Evans}, II, N.~J., {et~al.} 2007,
  \apjs, 171, 447

\bibitem[{{Reipurth} \& {Eiroa}(1992)}]{Reipurth}
{Reipurth}, B. \& {Eiroa}, C. 1992, \aap, 256, L1

\bibitem[{{Richer} {et~al.}(2000){Richer}, {Shepherd}, {Cabrit}, {Bachiller},
  \& {Churchwell}}]{Richer}
{Richer}, J.~S., {Shepherd}, D.~S., {Cabrit}, S., {Bachiller}, R., \&
  {Churchwell}, E. 2000, Protostars and Planets IV, 867

\bibitem[{{Rodriguez} {et~al.}(1989){Rodriguez}, {Curiel}, {Moran}, {Mirabel},
  {Roth}, \& {Garay}}]{Rodriguez}
{Rodriguez}, L.~F., {Curiel}, S., {Moran}, J.~M., {et~al.} 1989, \apjl, 346,
  L85

\bibitem[{{Smith} {et~al.}(2008){Smith}, {Buckle}, {Hills}, {Bell}, {Richer},
  {Curtis}, {Withington}, {Leech}, {Williamson}, {Dent}, {Hastings}, {Redman},
  {Wooff}, {Yeung}, {Friberg}, {Walther}, {Kackley}, {Jenness}, {Tilanus},
  {Dempsey}, {Kroug}, {Zijlstra}, \& {Klapwijk}}]{Smith}
{Smith}, H., {Buckle}, J., {Hills}, R., {et~al.} 2008, in Presented at the
  Society of Photo-Optical Instrumentation Engineers (SPIE) Conference, Vol.
  7020, Society of Photo-Optical Instrumentation Engineers (SPIE) Conference
  Series

\bibitem[{{Stanke} {et~al.}(2006){Stanke}, {Smith}, {Gredel}, \&
  {Khanzadyan}}]{Stanke}
{Stanke}, T., {Smith}, M.~D., {Gredel}, R., \& {Khanzadyan}, T. 2006, \aap,
  447, 609

\bibitem[{{Strom} {et~al.}(1976){Strom}, {Vrba}, \& {Strom}}]{Strom}
{Strom}, S.~E., {Vrba}, F.~J., \& {Strom}, K.~M. 1976, \aj, 81, 638

\bibitem[{{Sugitani} {et~al.}(2010){Sugitani}, {Nakamura}, {Tamura},
  {Watanabe}, {Kandori}, {Nishiyama}, {Kusakabe}, {Hashimoto}, {Nagata}, \&
  {Sato}}]{Sugitani}
{Sugitani}, K., {Nakamura}, F., {Tamura}, M., {et~al.} 2010, \apj, 716, 299

\bibitem[{{Testi} \& {Sargent}(1998)}]{Testi1}
{Testi}, L. \& {Sargent}, A.~I. 1998, \apjl, 508, L91

\bibitem[{{Testi} {et~al.}(2000){Testi}, {Sargent}, {Olmi}, \&
  {Onello}}]{Testi2}
{Testi}, L., {Sargent}, A.~I., {Olmi}, L., \& {Onello}, J.~S. 2000, \apjl, 540,
  L53

\bibitem[{{Torrelles} {et~al.}(1992){Torrelles}, {G{\'o}mez}, {Curiel},
  {Eiroa}, {Rodr{\'{\i}}guez}, \& {Ho}}]{Torrelles}
{Torrelles}, J.~M., {G{\'o}mez}, J.~F., {Curiel}, S., {et~al.} 1992, \apjl,
  384, L59

\bibitem[{{White} {et~al.}(1995){White}, {Casali}, \& {Eiroa}}]{White}
{White}, G.~J., {Casali}, M.~M., \& {Eiroa}, C. 1995, \aap, 298, 594

\bibitem[{{Williams} \& {Myers}(2000)}]{Williams}
{Williams}, J.~P. \& {Myers}, P.~C. 2000, \apj, 537, 891

\bibitem[{{Winston} {et~al.}(2007){Winston}, {Megeath}, {Wolk}, {Muzerolle},
  {Gutermuth}, {Hora}, {Allen}, {Spitzbart}, {Myers}, \& {Fazio}}]{Winston}
{Winston}, E., {Megeath}, S.~T., {Wolk}, S.~J., {et~al.} 2007, \apj, 669, 493

\bibitem[{{Wolf-Chase} {et~al.}(1998){Wolf-Chase}, {Barsony}, {Wootten},
  {Ward-Thompson}, {Lowrance}, {Kastner}, \& {McMullin}}]{Wolf-Chase}
{Wolf-Chase}, G.~A., {Barsony}, M., {Wootten}, H.~A., {et~al.} 1998, \apjl,
  501, L193+

\bibitem[{{Wu} {et~al.}(2004){Wu}, {Wei}, {Zhao}, {Shi}, {Yu}, {Qin}, \&
  {Huang}}]{Wu}
{Wu}, Y., {Wei}, Y., {Zhao}, M., {et~al.} 2004, \aap, 426, 503

\bibitem[{{Ziener} \& {Eisl{\"o}ffel}(1999)}]{Ziener}
{Ziener}, R. \& {Eisl{\"o}ffel}, J. 1999, \aap, 347, 565

\end{thebibliography}

\clearpage


\clearpage

\clearpage






\clearpage





\clearpage



\end{document}